\documentclass[%
 reprint,
nofootinbib,
 amsmath,amssymb,
 aps,
]{revtex4-2}

\usepackage{graphicx}
\usepackage{dcolumn}
\usepackage{bm}
\usepackage{subcaption}
\usepackage{color}

\begin{document}

\preprint{APS/123-QED}

\title{Fragility of spectral clustering for networks with an overlapping structure}

\author{Chihiro Noguchi}
 \email{noguchi.c.aa@m.titech.ac.jp}
 \affiliation{Department of Mathematical and Computing Science, Tokyo Institute of Technology, 2-12-1 Ookayama, Meguro-ku, Tokyo, Japan}
\author{Tatsuro Kawamoto}%
\affiliation{%
    Artificial Intelligence Research Center, National Institute of Advanced Industrial Science and Technology, 2-3-26 Aomi, Koto-ku, Tokyo, Japan
}%

\date{\today}

\begin{abstract}
Overlapping communities are commonly observed in real-world networks. This is a motivation to develop overlapping community detection methods, because methods for non-overlapping communities may not perform well. However, deterioration mechanism of the detection methods used for non-overlapping communities have rarely been investigated theoretically. Here, we analyze accuracy of spectral clustering, which does not consider overlapping structures, by using the replica method from statistical physics. Our analysis on an overlapping stochastic block model reveals how the structural information is lost from the leading eigenvector because of the overlapping structure. 
\end{abstract}

\maketitle

\begin{figure*}[t!]
\centering
\begin{tabular}{c}

    \begin{minipage}{0.45\hsize}
    \includegraphics[clip, width=6cm]{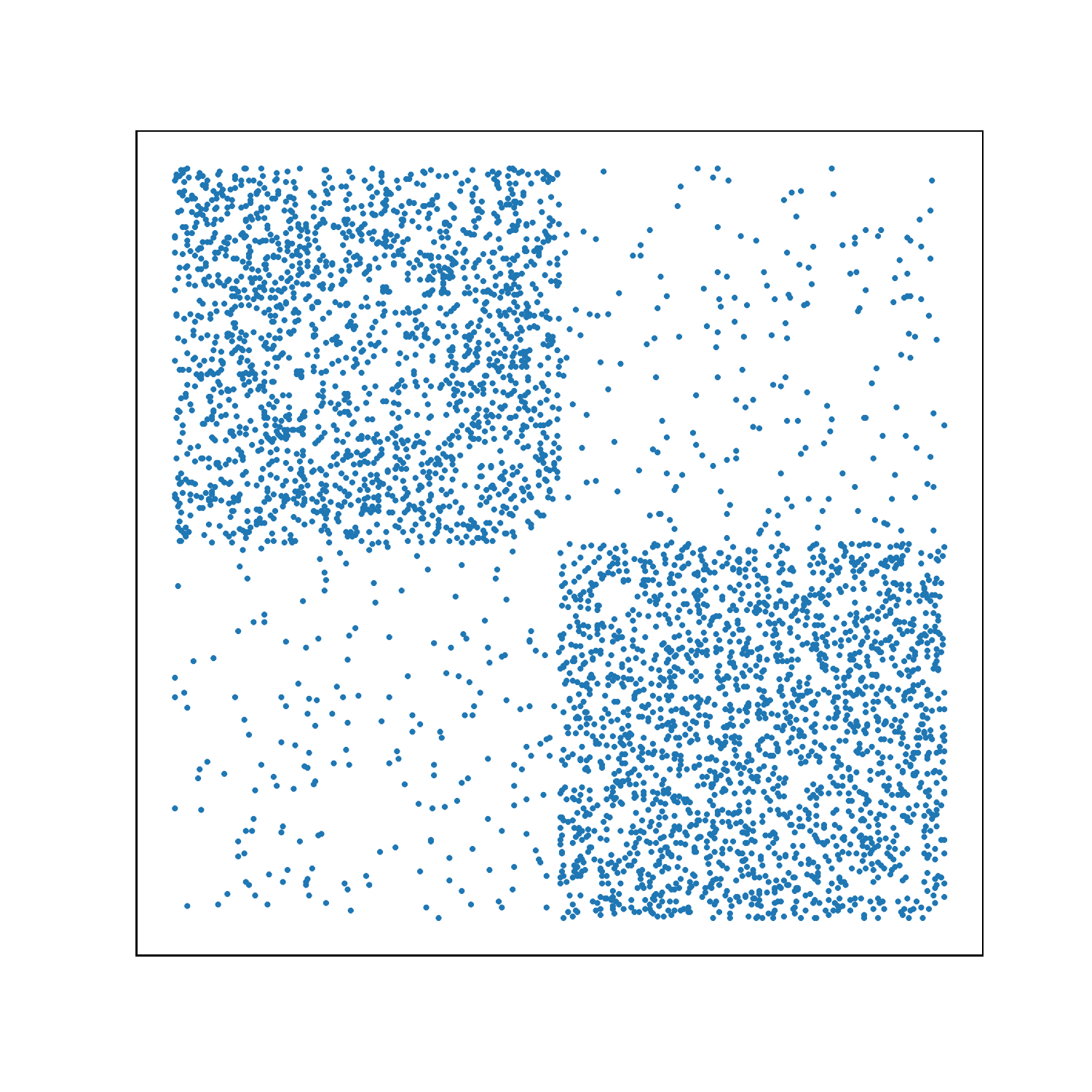}
    \vspace{-0.5cm}
    \subcaption{}
    \label{fig:standard_sbm_instances1}
    \end{minipage}
    
    \begin{minipage}{0.45\hsize}
    \includegraphics[clip,width=6cm]{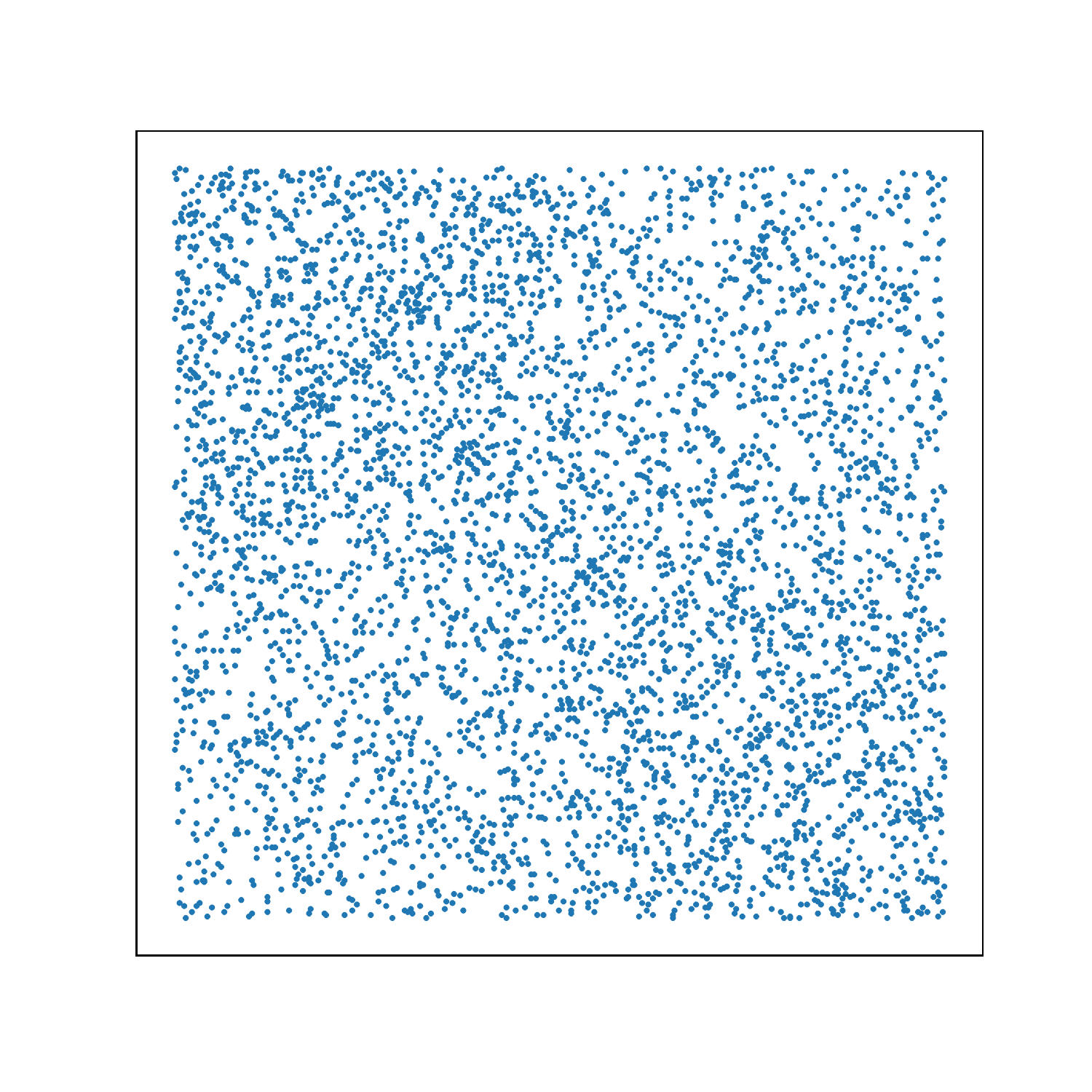}
    \vspace{-0.5cm}
    \subcaption{}
    \label{fig:standard_sbm_instances2}
    \end{minipage}
    
    \vspace{0.5cm}
    \hspace{-0.9cm}
    
    \\
    
    \begin{minipage}{0.45\hsize}
    \includegraphics[clip, width=6cm]{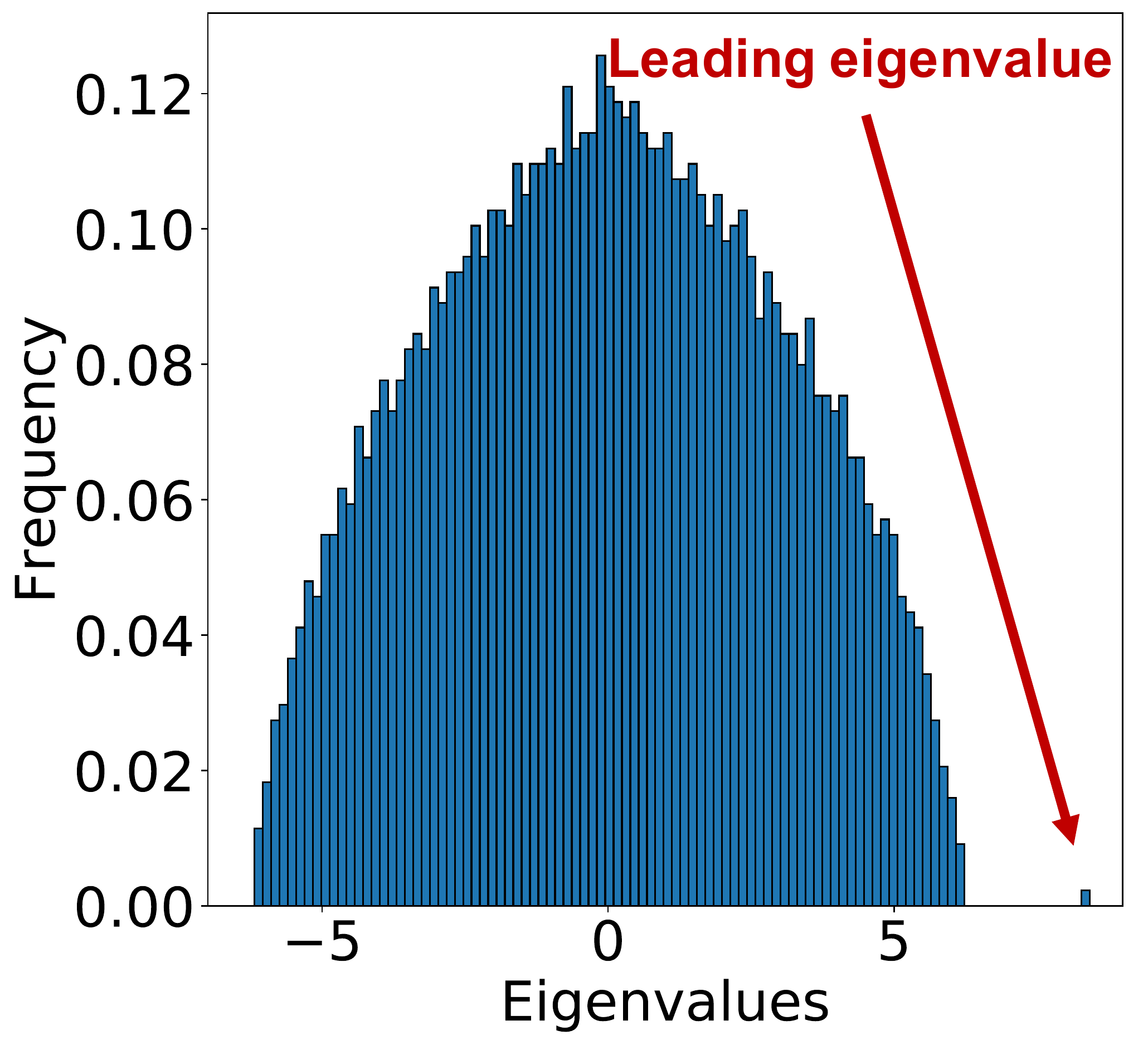}
    \subcaption{}
    \label{fig:bulk_eigenvalues1}
    \end{minipage}
    
    \begin{minipage}{0.45\hsize}
    \includegraphics[clip,width=6cm]{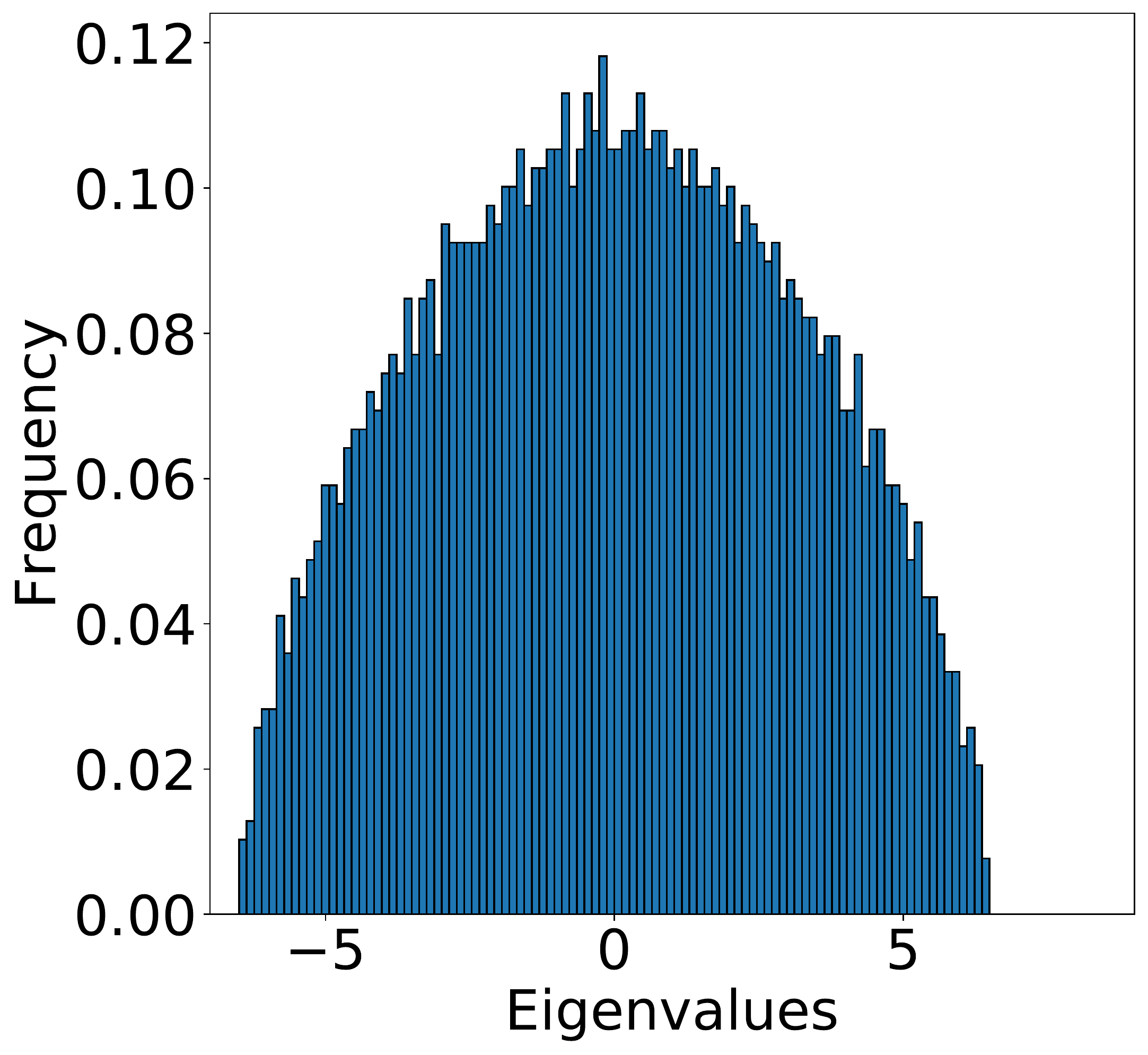}
    \subcaption{}
    \label{fig:bulk_eigenvalues2}
    \end{minipage}
    
\end{tabular}
\caption{Adjacency matrices of graphs with a non-overlapping structure and the corresponding histograms of the bulk of eigenvalues of the modularity matrix. (a, c) Nodes in the same communities are more densely connected internally than externally (strong community structure). (b, d) All nodes are connected with almost the same probability (weak community structure).}
\label{fig:standard_sbm_instances}
\end{figure*}

\section{Introduction}
A graph or a network that represents related data is a common data structure in multivariate statistics, machine learning, and statistical mechanics. Identifying densely connected subgraphs---community detection---is useful for graph analysis. Such subgraphs (or the corresponding node set) are referred to as communities. 
Spectral clustering is a popular community detection algorithm that is efficient yet highly accurate on random graph models \cite{nadakuditi2012graph,krzakala2013spectral,mossel2014belief,abbe2017entrywise}.
Nevertheless, spectral clustering often fails to identify plausible communities when it is applied to real-world networks. This is presumably because of specific features of real-world networks that are missing in simple random graph models. To fill this discrepancy, in this paper, we theoretically investigate how overlapping of communities affects accuracy of spectral clustering. 
We will give precise definitions of a community, an overlapping community, and the accuracy of clustering in Sec.~\ref{sec:overlapping_sbm}.

We denote an undirected graph as $G=(V,E)$, where $V$ ($|V|=N$) is a set of nodes and $E$ ($|E|=m$) is a set of edges. The graph is represented by the $N \times N$ adjacency matrix $A$, where $A_{ij} = 1$ when a pair of nodes $i$ and $j$ is connected by an edge and $A_{ij} = 0$ otherwise. The adjacency matrix of graphs with strong (Fig.~\ref{fig:standard_sbm_instances1}) and weak (Fig.~\ref{fig:standard_sbm_instances2}) non-overlapping community structures are illustrated in Fig.~\ref{fig:standard_sbm_instances}.

To identify the community structure, spectral clustering \cite{von2007tutorial} computes the leading eigenvalues and eigenvectors of a regularized adjacency matrix; in this paper, as an example, we focus on the so-called modularity matrix \cite{newman2006modularity} as the regularized adjacency matrix. When the community structure can be clearly identified, the isolated leading eigenvectors have relevant information of the communities, while a bulk of eigenvalues emerges from the randomness of a graph. For example, Fig.~\ref{fig:bulk_eigenvalues1} shows the spectral density of the modularity matrix corresponding to the adjacency matrix in Fig.~\ref{fig:standard_sbm_instances1}. In this case, the largest eigenvalue is clearly separated from the bulk of eigenvalues, and we can extract two communities using the isolated leading eigenvector. On the other hand, Fig.~\ref{fig:bulk_eigenvalues2} shows the case corresponding to the adjacency matrix in Fig.~\ref{fig:standard_sbm_instances2}. The eigenvalue correlated to the community structure is buried in the bulk of eigenvalues, and the spectral density is no longer distinguishable from that of a uniformly random graph. The phase transition point that the eigenvalues do not exhibit community structure at all is referred to as the (algorithmic) detectability limit \cite{nadakuditi2012graph,kawamoto2015limitations,kawamoto2015detectability} of spectral clustering. 

As a tool for theoretical analysis, we use the replica method that originated from statistical physics. It enables us to calculate the ensemble average over random graph instances. As a result, we obtain a detectability phase diagram that indicates the effect of overlapping on spectral clustering. 

Several existing studies have investigated the fragility, i.e., lack of robustness, of spectral clustering.
Owing to the fact that real-world networks have more complex structures than a simple random graph, the studies have considered the fragility in case of, e.g., adversarial perturbations \cite{stephan2018robustness}, noise perturbations \cite{li2007noise,balakrishnan2011noise}, tangles and cliques \cite{abbe2018graph}, and localization of eigenvectors \cite{kawamoto2015limitations,kawamoto2015detectability,zhang2016robust}.
In this paper, we analyze the effect of the overlapping structure on the graph spectra. Specifically, we found that, when the size of the community overlap is increased, it is the isolated eigenvalue that is mainly affected. On the other hand, it is the bulk of eigenvalues that is mainly affected when the density of the community overlap is increased.

Noted that identifying an overlapping community structure itself is not a goal of this paper. There are in fact many algorithms for such a purpose \cite{palla2005uncovering,ahn2010link,benson2016higher,peixoto2015model,airoldi2008mixed,wang2011community,psorakis2011overlapping,yang2013overlapping,rosvall2014memory}. To identify or to assess an overlapping community structure, one should use a suitable algorithm. Usually, however, we do not a priori know whether communities are overlapped. Moreover, even when it is the case, it is hard to imagine that the spectral clustering becomes completely useless. Thus, we investigate how the signal of structural heterogeneity remains in the spectral clustering, although it may not be the best algorithm to use.

The rest of the paper is organized as follows. In Sec.~\ref{sec:overlapping_sbm}, we introduce the overlapping random graph models that we consider. In Sec.~\ref{sec:replica_analysis}, we provide the replica analysis for the graph spectra of the random graph model. In Sec.~\ref{sec:numerical_experiments}, we show the results and their interpretation obtained by the replica analysis. Finally, Sec.~\ref{sec:summary} presents a discussion.

\begin{figure*}[t!]
\centering
\hspace{-3cm}
\begin{tabular}{c}

\hspace{-2cm}
\begin{minipage}{0.4\hsize}
\centering
\includegraphics[clip, width=12cm]{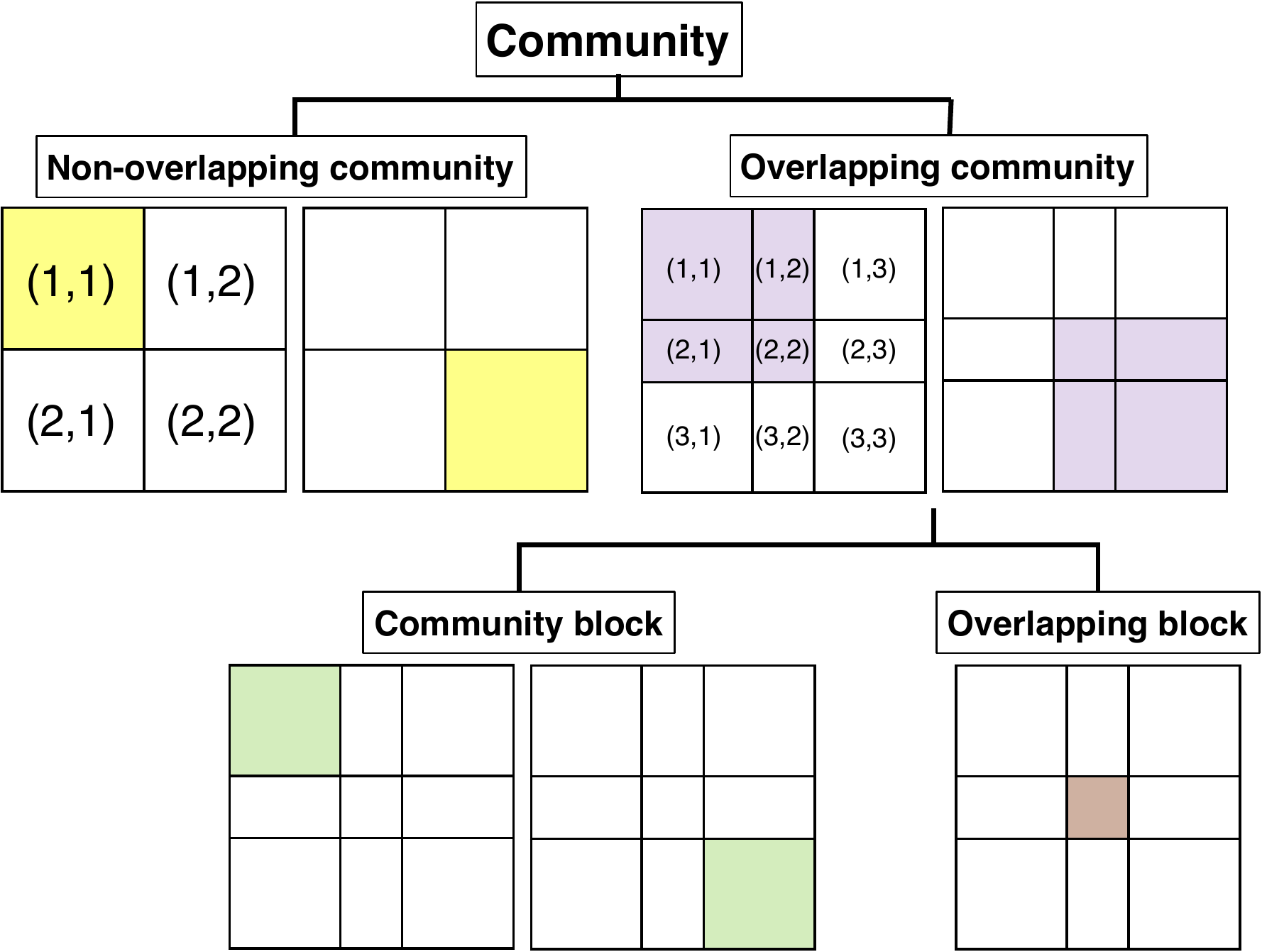}
\subcaption{}
\label{fig:communities}
\end{minipage}

\\

\begin{minipage}{0.3\hsize}
\centering
\includegraphics[clip, width=5cm]{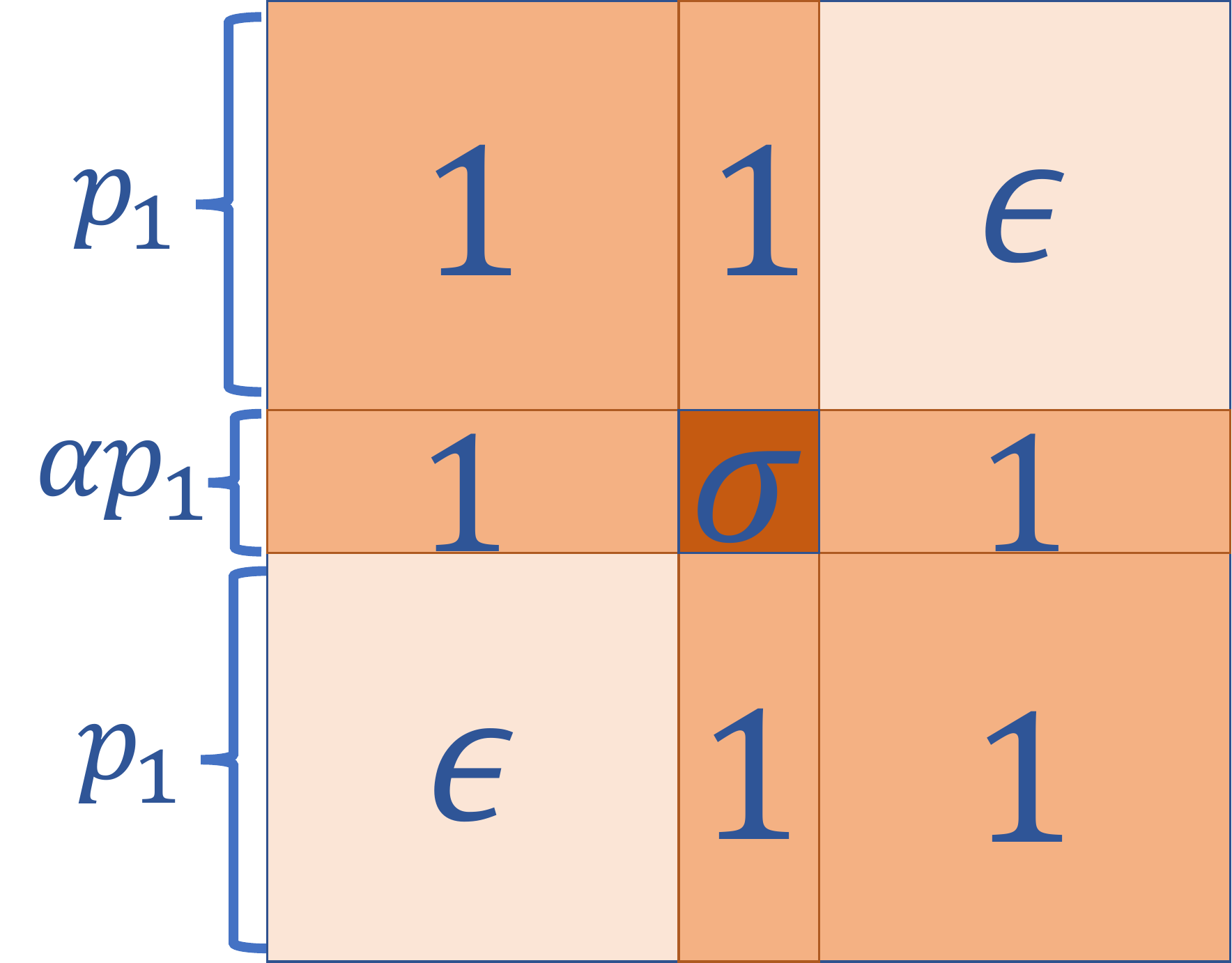}
\subcaption{}
\label{fig:overlapping_sbm}
\end{minipage}

\begin{minipage}{0.4\hsize}
\includegraphics[clip, width=10cm]{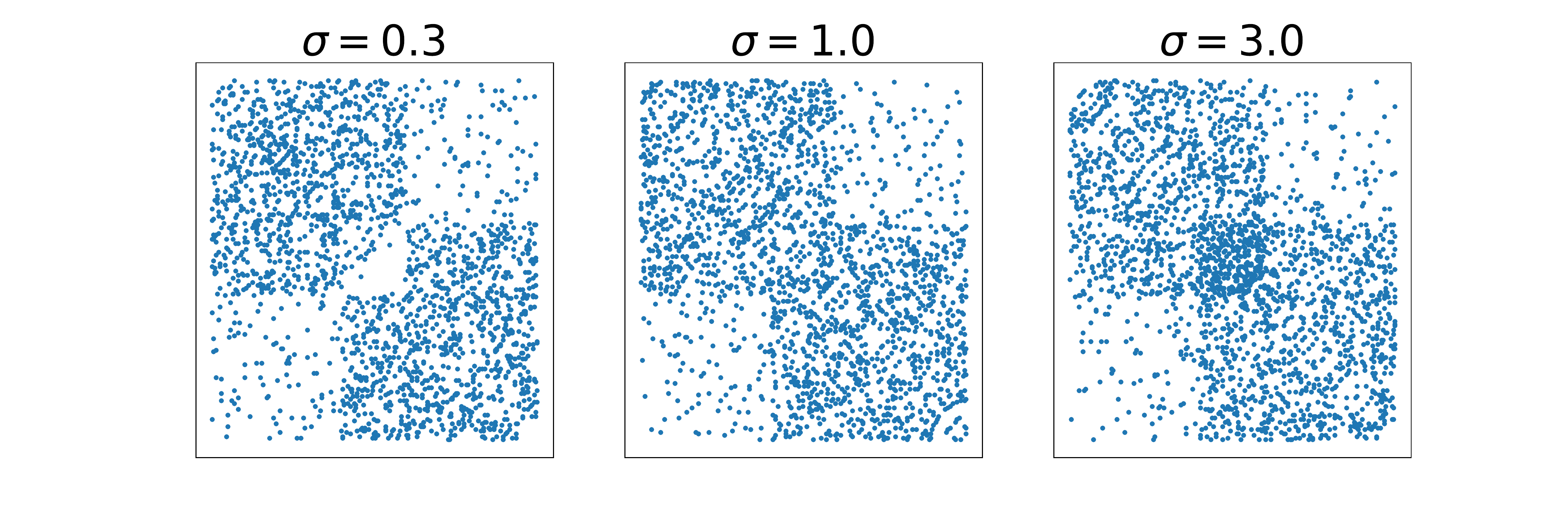}
\subcaption{}
\label{fig:sbm_graph_instance}
\end{minipage}


\end{tabular}

\caption{(a) Classification chart of communities and blocks using schematic pictures of adjacency matrices. Communities are classified into non-overlapping and overlapping communities, and an overlapping community consists of nodes in community blocks and nodes in an overlapping block. Each matrix element represents a pair of block labels, i.e., the corresponding set of node pairs. (b) Structure of the overlapping SBM that we consider. The node sets incident on the $(1,1)$ and $(3,3)$ elements correspond to the community blocks, while the node set incident on the $(2,2)$ element corresponds to the overlapping block. $\alpha$, $\epsilon$, and $\sigma$ are the fraction of the overlapping block size, the inverse of the community structure strength, and the density of the overlapping block, respectively. The value of each block corresponds to an element of affinity matrix (\ref{eq:can_affinity_matrix}) divided by $\rho_{\operatorname{in}}$. (c) Adjacency matrices of graph instances of the overlapping SBM with $\sigma=0.3, 1 $, and $3$.}
\label{fig:overlapping_model_example}
\end{figure*}

\section{Overlapping stochastic block model}
\label{sec:overlapping_sbm}
Throughout this paper, we consider a class of random graph models called the stochastic block model (SBM). It is a random graph model that has a preassigned (planted) modular structure. Here, as a particular case of the SBM, we introduce the overlapping SBM. Although we focus only on the so-called canonical SBM in the main text, its microcanonical counterpart \cite{peixoto2017bayesian,peixoto2017nonparametric} (see Appendix~\ref{sec:micro_stochas_block_model} for a detailed definition) is also analyzed in Appendix ~\ref{sec:appendix_derivation_micro}.

\subsection{Canonical SBM}
\label{sec:can_stochas_block_model}
Before considering the overlapping SBM, we first introduce the (canonical) SBM with a general structure. 
We define a \textit{block} as a node set in which the nodes are statistically equivalent.
A graph with $K$ blocks is generated from the SBM as follows. For each node of the graph, we preassigned a block label $\boldsymbol{t}=[t_i]\ t_i\in\{1,\cdots,K\} \ (i\in V)$. Then, each pair of nodes $(i,j)$ is connected by an edge with probability $\rho_{t_it_j}$ independently and randomly; this probability is provided as an element of the $K\times K$ affinity matrix $\boldsymbol{\rho}=[\rho_{kl}],\ 0\le\rho_{kl}\le 1$.
Therefore, the probability of a graph instance is expressed as
\begin{equation}
    P(A|K,\boldsymbol{t},\boldsymbol{\rho})=
    \prod_{i<j}\rho_{t_it_j}^{A_{ij}}(1-\rho_{t_it_j})^{1-A_{ij}}.
    \label{eq:canonical_sbm_probability}
\end{equation}
Here, because we consider undirected simple graphs, we assume that $A_{ii}=0$ and $A_{ij}=A_{ji}$. Moreover, we focus on sparse graphs throughout this paper; i.e., we assume $\rho_{rs}=O(1/N)$ for all $r$ and $s$. When every matrix element of $\boldsymbol{\rho}$ is equal, the model becomes the so-called Erd\H{o}s--R\'{e}nyi random graph model.
We also introduce a vector that represents the block-size distribution as $\boldsymbol{p}=[p_k]\ (k\in\{1,\dots,K\})$, where $p_{k} = \sum_{i=1}^{N} \delta_{k, t_{i}}/N$ ($\delta_{a,b}$ represents Kronecker's delta).

For example, a two-block SBM is parameterized as
\begin{align}
    \boldsymbol{p}&=(p_1,p_2),\label{eq:non_overlap_can_group_sizes} \\
    \boldsymbol{\rho} &=
    \begin{pmatrix}
    \rho_{\text{in}} & \rho_{\text{out}} \\
    \rho_{\text{out}} & \rho_{\text{in}}
    \end{pmatrix}
    =
    \begin{pmatrix}
    1 & \epsilon \\
    \epsilon & 1
    \end{pmatrix}\rho_{\operatorname{in}}.\label{eq:non_overlap_can_affinity_matrix}
\end{align}
Here, edges in the $(1,1)$ and $(2,2)$ elements have the same generation probability $\rho_{\operatorname{in}}$.
In contrast, edges in the $(1,2)$ and $(2,1)$ elements have the generation probability $\rho_{\operatorname{out}}$; $\epsilon=\rho_{\operatorname{out}}/\rho_{\operatorname{in}}$ is a parameter that controls the strength of the community structure.
We define \textit{non-overlapping communities} as node sets incident on the $(1,1)$ and $(2,2)$ elements, as illustrated in Fig.~\ref{fig:communities}.

\subsection{Overlapping canonical SBM}

We define the overlapping SBM as the three-block SBM that has parameters
\begin{align}
    \boldsymbol{p}&=(p_1,p_2,p_3),\label{eq:can_group_sizes} \\
    \boldsymbol{\rho} &=
    \begin{pmatrix}
    \rho_{\text{in}} & \rho_{\text{in}} & \rho_{\text{out}} \\
    \rho_{\text{in}} & \sigma
    \rho_{\text{in}} & \rho_{\text{in}} \\
    \rho_{\text{out}} & \rho_{\text{in}} & \rho_{\text{in}}
    \end{pmatrix}
    =
    \begin{pmatrix}
    1 & 1 & \epsilon \\
    1 & \sigma & 1 \\
    \epsilon & 1 & 1
    \end{pmatrix}\rho_{\operatorname{in}}.\label{eq:can_affinity_matrix}
\end{align}
Here, $\boldsymbol{p}$ and $\boldsymbol{\rho}$ are illustrated in Fig.~\ref{fig:overlapping_sbm}.
As illustrated in Fig.~\ref{fig:communities}, we define the node sets incident on the sets of elements $\{(1,1)$, $(1,2)$, $(2,1)$, $(2,2)\}$ and $\{(2,2)$, $(2,3)$, $(3,2)$, $(3,3)\}$ as \textit{overlapping communities}, respectively; edges therein have the same generation probability $\rho_{\mathrm{in}}$, except for the $(2,2)$ element. 
Within the overlapping communities, we define the node sets incident on the $(1,1)$ and $(3,3)$ elements as \textit{community blocks} and the node set incident on the $(2,2)$ element as an \textit{overlapping block}.
We let the edge generation probability of the $(1,3)$ and $(3,1)$ elements be $\rho_{\mathrm{out}}$ ($=\epsilon\rho_{\operatorname{in}}$). The edge generation probability of the $(2,2)$ element is parametrized as $\sigma\rho_{\mathrm{in}}$; $\sigma$ is a parameter that controls the density of the overlapping block.
Adjacency matrices with different values of $\sigma$ are exemplified in Fig.~\ref{fig:sbm_graph_instance}
(see Appendix~\ref{sec:relationship_with_MMSBM} for the relationship between this overlapping SBM and the mixed-membership SBM \cite{airoldi2008mixed}.)

We define the average degree of each block $\boldsymbol{c}=(c_1, c_2, c_3)$, where the degree of a node is the number of edges connected to the node.
The ratio $c_1/c_2$ can also be expressed as $(1+\alpha+\epsilon)/(\sigma\alpha+2)$ using the affinity matrix elements, where we introduced $\alpha\equiv p_2/p_1$. Therefore, the parameters of the overlapping SBM are constrained as
\begin{equation}
    c_1(\sigma\alpha+2)=c_2(1+\alpha+\epsilon).
    \label{eq:constraint_between_alp_eps}
\end{equation}
For simplicity, we assume the symmetry between the community blocks, i.e., $p_1=p_3$ and $c_1=c_3$. We assume that the affinity matrix is symmetric, owing to the fact that we consider undirected graphs.

A technically interesting aspect of the present analysis is that this is a model-inconsistent scenario; while the overlapping SBM that we consider consists of three blocks, we consider the partitioning into two non-overlapping communities.

How to evaluate the accuracy of the spectral clustering on the overlapping SBM is an arguable issue. In this paper, we evaluate whether the community blocks are identified correctly and neglect the partitioning with respect to the overlapping block. That is, we define an accuracy of a partition as 
\begin{align}
    &\operatorname{Accuracy}\equiv\max\left\{f(\boldsymbol{\hat{t}}),f\left(\mathcal{P}(\boldsymbol{\hat{t}})\right)\right\},\label{eq:model_inconsistent_accuracy}\nonumber \\
    &f(\boldsymbol{\hat{t}})=
    \frac{1}{N(p_1+p_3)}
    \left(\sum_{i\in V_1} \delta_{\hat{t}_i, 1}
    +
    \sum_{i\in V_3} \delta_{\hat{t}_i, 2}\right).
\end{align} 
Here, $\sum_{i\in V_k}$ is the sum over the node indices belonging to the $k$th block, $\boldsymbol{\hat{t}}=[\hat{t}_i]\ (\hat{t}_i\in\{1,2\})$ is the inferred non-overlapping community label of node $i$. 
The operator $\mathcal{P}$ permutes the inferred labels; namely, $\hat{t}_i=1$ is replaced by $\hat{t}_i=2$ and vice versa.
The maximization is required to eliminate the degrees of freedom by permutation.

\section{Replica analysis}
\label{sec:replica_analysis}
We now calculate the spectrum of the overlapping SBM and show that a phase transition point of the largest eigenvalue exhibits the detectability limit. It should be noted that the same result is obtained in the case of the microcanonical SBM (Appendix~\ref{sec:appendix_derivation_micro}).

\subsection{Spectrum and the detectability limit of the overlapping SBM}
\label{sec:replica_analysis_for_canonical}
As an example of a regularized adjacency matrix, we consider the modularity matrix. Each element of the matrix is defined as
\begin{equation}
    M_{ij} = A_{ij} - \frac{d_id_j}{2m},
\end{equation}
where $d_i$ $(=\sum_{j=1}^NA_{ij})$ is the degree of a node $i$ and $m$ $(=|E|)$ is the total number of the edges. Partitioning into two non-overlapping communities can be identified by the eigenvector of the largest eigenvalue. Thus, our goal is to solve the following maximization problem.
\begin{equation}
    \lambda(M)=\frac{1}{N}\max _{\boldsymbol{x}}\,\boldsymbol{x}^{\top} M \boldsymbol{x}, \quad \text { subj. to } \boldsymbol{x}^{\top} \boldsymbol{x}=N,
    \label{eq:modularity_maximization_problem2}
\end{equation}
where $\lambda(M)$ is the largest eigenvalue of $M$, and $\boldsymbol{x}^{\top}$ is the transpose of a vector $\boldsymbol{x}$.
This problem can be expressed as
\begin{align}
    f(M,\beta) &= -\frac{1}{\beta N}\log Z(M, \beta),\label{eq:free_energy_log_z} \\
    \lambda(M) &= -2\lim_{\beta\to\infty} f(M,\beta), \label{eq:beta_inf_free_energy} \\
    Z(M,\beta) &= \int d\boldsymbol{x} e^{\frac{\beta}{2}\boldsymbol{x}^\top M\boldsymbol{x}}\delta(\boldsymbol{x}^\top\boldsymbol{x}-N),
    \label{eq:partition_function}
\end{align}
where $Z(M,\beta)$ is the partition function. The constraint (\ref{eq:modularity_maximization_problem2}) is imposed by the delta function in (\ref{eq:partition_function}), and taking $\beta\to\infty$ in (\ref{eq:beta_inf_free_energy}) leads to the maximization of the exponent of the exponential function in (\ref{eq:partition_function}).
Because we are interested in the typical behavior of the graph instances, we analyze
\begin{equation}
    [\lambda(M)]_M=2\lim_{\beta\to\infty}\frac{1}{\beta N}\left[\log Z(M,\beta)\right]_M,
    \label{eq:lambda_free_energy}
\end{equation}
where $[\cdots]_M$ represents the ensemble average over graph instances. Unfortunately, it is difficult to calculate the average $\left[\log Z(M,\beta)\right]_M$ analytically. To overcome this difficulty, we use the replica trick, namely,
\begin{equation}
    \left[\log Z(M,\beta)\right]_M=\lim_{n\to 0}\frac{\partial}{\partial n}\log [Z^n(M,\beta)]_M.
    \label{eq:replica_trick}
\end{equation}
Here, the exponent $n$ in $[Z^n]_M$ is a real value. However, we treat $n$ as an integer for a moment. In the end, we perform the analytic continuation to the real value.
This treatment is termed the replica method.

From Eq.~(\ref{eq:partition_function}), the $n$th moment the partition function is obtained as
\begin{align}
    &[Z^n(M,\beta)]_M \nonumber \\
    &=
    \int \left(\prod_{a=1}^nd\boldsymbol{x}_a\delta(\boldsymbol{x}_a^\top\boldsymbol{x}_a-N)\right)
    \left[
    \exp\left(\frac{\beta}{2}\sum_a\boldsymbol{x}_a^\top M\boldsymbol{x}_a\right)\right]_M,
    \label{eq:can_nth_moment_Z}
\end{align}
where $a\in \{1,\dots,n\}$ is an index of $n$ identical copies.
For further calculations, we introduce several order parameters and approximations. Detailed calculations are described in Appendix~\ref{sec:appendix_b}. As a result, the average largest eigenvalue in the limit of $N\to\infty$ is obtained by the following saddle-point (extremum) condition of nine auxiliary variables $(\phi, \Omega, \hat{\Omega}, m_{1k}, m_{2k},\hat{m}_{1k}, \hat{m}_{2k},a_k,\hat{a}_k)$.
\begin{widetext}
\begin{align}
    \left[ \lambda(M) \right] _ { M } 
    &= \underset{\phi, \Omega, \hat{\Omega}, m_{1k}, m_{2k},\hat{m}_{1k}, \hat{m}_{2k}, a_k,\hat{a}_k}{\operatorname { extr }} \left\{ \phi + 2 \hat { \Omega } \Omega - \Omega ^ { 2 }\right. \nonumber \nonumber\\
   &\ + \frac { 1 } { 2 }N \sum _ { k , k ^ { \prime } } W _ { k k ^ { \prime } } \left( \frac { a_{k'} 
   \left(m _ { 2 k }-\frac{2\hat{\Omega}}{\sqrt{\bar{c}}}+\frac{4\hat{\Omega}^2}{\bar{c}}\right)
   + 2 m _ { 1 k' }
   \left(m_{1k}-\frac{2\hat{\Omega}}{\sqrt{\bar{c}}}\right)
   + a_{k} m _ { 2 k ^ { \prime } } } { a_k a_{k'} - 1 }
   -\frac{m_{2k}-\frac{2\hat{\Omega}}{\sqrt{\bar{c}}}m_{1k}+\frac{4\hat{\Omega}^2}{c}}{a_k}-\frac{m_{2k'}}{a_{k'}}\right) \nonumber\\
    &\ - \sum _ { k } p _ { k } c _ { k } \left( \frac { m _ { 2 k } + 2 m _ { 1 k } \hat { m } _ { 1 k } + \hat { m } _ { 2 k } } { a_k - \hat { a }_k } -\frac{m_{2k'}}{a_{k'}} \right) \nonumber\\
    &\ + \left.\frac { 1 } { N } \sum _ { k } \sum _ { i \in V _ { k } } \sum _ { d = 0 } ^ { \infty } \frac { \mathcal { P } _ { c _ { k } } ( d ) } { \phi - d \hat { a }_k } \left( d \hat { m } _ { 2 k } + d ( d - 1 ) \hat { m } _ { 1 k } ^ { 2 } \right) \right\}.
    \label{eq:can_lam}
\end{align}
\end{widetext}
Here, $W_{kl}$ and $\bar{c}$ are defined as $W_{kl}\equiv p_k\rho_{kl}p_l$ and $\bar{c}\equiv 2m/N$, respectively. $\mathcal{P}_{c_k}(d)$ is the Poisson probability mass function of degree $d$ of each node in block $k$ that has expectation $c_k$.
$m_{1k}$ is the the mean of the largest eigenvector elements that corresponds to the $k$th block.
$m_{1k}$ plays an important role in the derivation of the detectability limit. Definitions and interpretations of the other auxiliary variables are omitted here, because they are not directly relevant to the detectability limit (see Appendix~\ref{sec:appendix_b} for the precise definitions).

The detectability limit is derived by solving the equations of the nine auxiliary variables. In particular, $m_{11}$ $(=-m_{13})$ plays an important role for the detectability limit. When $m_{11}^2>0$, the spectral clustering retains the ability to detect the community structure better than a random guess (detectable condition). On the other hand, when $m_{11}^2=0$, the result of spectral clustering is uncorrelated to the planted structure (undetectable condition). Accordingly, the phase transition point is derived by the condition $m_{11}^2=0$. 
This corresponds to the condition that the largest eigenvalue is buried in the bulk of the eigenvalues, as we mentioned in Introduction.

\section{Accuracy of the spectral clustering on the overlapping SBM}
\label{sec:numerical_experiments}
In this section, using the results obtained by the replica analysis, we show how the size and density of the overlapping block affect the spectrum.
We also check the validity of our analytical calculations by comparing them to the results of numerical experiments.
Here, we use the microcanonical SBM in the numerical experiments instead of the canonical SBM.
Here, for a technical reason that we describe in Appendix~\ref{sec:appendix_comparison_two_models}, we use the microcanonical counterpart of the SBM.
We used graph-tool \cite{peixoto2014graph} to generate graph instances of the microcanonical SBM.

\subsection{Detectability phase diagram and the leading eigenvalue}
\label{sec:detectability_phase_diagram_and_the_lleading_eigenvalue}

\begin{figure}[t!]
\centering
\includegraphics[width=9cm]{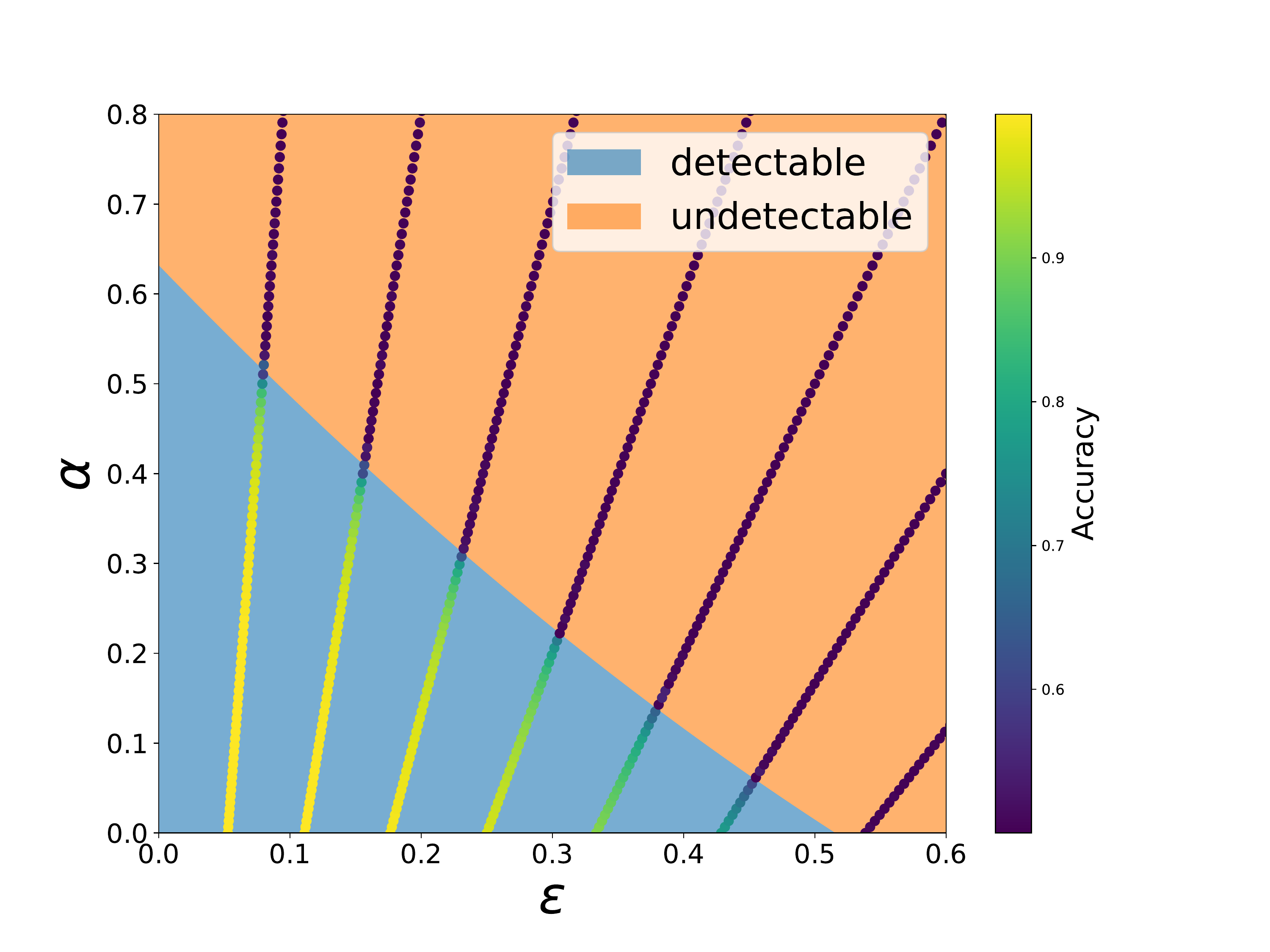}
\caption{Detectability phase diagram of the $(\epsilon, \alpha)$ plane. The model parameters are set to $c_1=10$ and $\sigma=2$. Along the line of the results of the numerical experiments determined by constraint (\ref{eq:constraint_between_alp_eps}), degree $c_2$ takes a fixed value. The lines in this figure, from left to right, correspond to the values of $c_2$ from 19 to 11.}
\label{fig:phase_diagram}
\end{figure}

\begin{figure}[t!]
\centering
\includegraphics[clip, width=9cm]{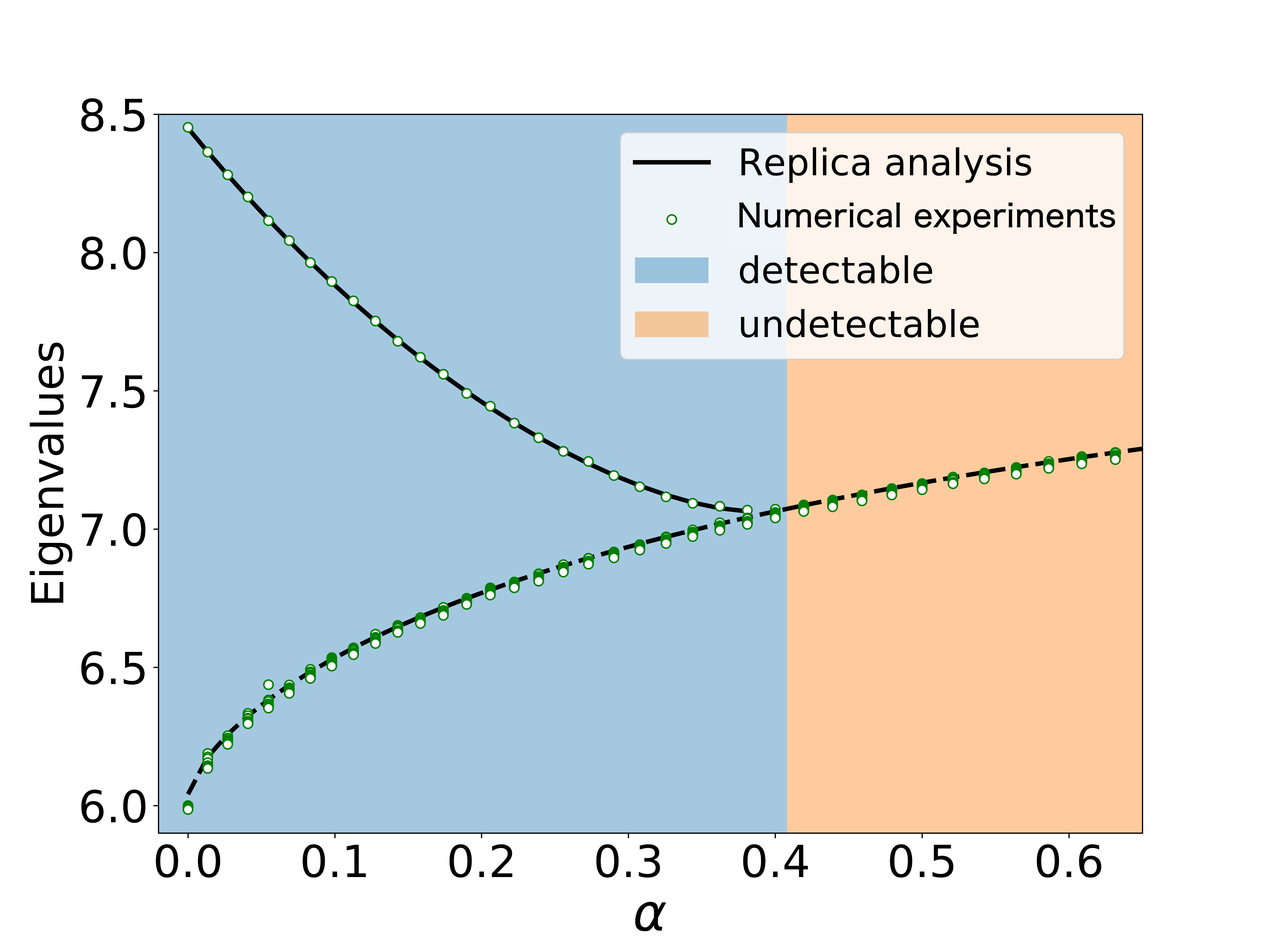}
\caption{Eigenvalues derived by the replica analysis as a function of $\alpha$. We set $c_1=10$, $c_2=18$, and $\sigma=2$.
The solid and dashed lines represent the isolated leading eigenvalues and the bulk edges of eigenvalues, respectively. The boundary between the blue and orange regions represents the detectability limit. The green dots represent the top ten eigenvalues computed in the numerical experiments.}
\label{fig:top10_eigenvalues_overlap}
\end{figure}

First, to observe the overall dependency of overlapping structures, we show the detectability phase diagram.
Figure~\ref{fig:phase_diagram} shows the detectability phase diagram of the $(\epsilon,\alpha)$ plane. As mentioned above, $\epsilon$ is the parameter that controls the strength of the community structure and $\alpha=p_2/p_1$ is the ratio of the overlapping block and a community block. The boundary between the blue and orange regions represents the detectability limit of the spectral clustering predicted by the replica analysis. The dots represent the results of the numerical experiments; the color gradient represents the accuracy defined in Eq.~(\ref{eq:model_inconsistent_accuracy}).
We can see that both boundaries are in a good agreement.
Note that the numerical experiment is possible only on specific curves in the parameter space because of constraint (\ref{eq:constraint_between_alp_eps}), and $c_2$ can take only natural numbers in the microcanonical SBM.
In this experiment, we set $c_1=10$ and $\sigma=2$. Then, the range $c_2$ can take is restricted between 11 and 19 because of the assortative condition $0\leq\epsilon\leq1$.
This phase diagram is the result that shows how fragile the spectral clustering is against the overlapping structure. 

Figure~\ref{fig:top10_eigenvalues_overlap} shows the leading eigenvalue and the edge of the bulk of the eigenvalues\footnote{The edge of the bulk of eigenvalue is derived as the largest eigenvalue under the undetectable condition.}, which are predicted by the replica analysis, and the top ten eigenvalues computed in the numerical experiments. We can confirm that the replica analysis accurately describes the behavior of numerical experiments.
When $\alpha$ is small, the leading eigenvalue is separated from the bulk of the eigenvalues. As $\alpha$ increases, the leading eigenvalue approaches the bulk of the eigenvalues. As we described in Introduction, when it reaches the bulk of the eigenvalues, the spectral clustering loses ability to detect the community structure, i.e., the detectability limit. Note the value of $\epsilon$ also varies according to (\ref{eq:constraint_between_alp_eps}) as $\alpha$ varies. Thus, the horizontal axis in Fig.~\ref{fig:top10_eigenvalues_overlap} corresponds to the line in Fig.~\ref{fig:phase_diagram} with $c_2=18$.

\subsection{Effects of the size of the overlapping structure}
\label{sec:comparison_with_bimodal}

\begin{figure}[t!]
\centering
\includegraphics[width=9cm]{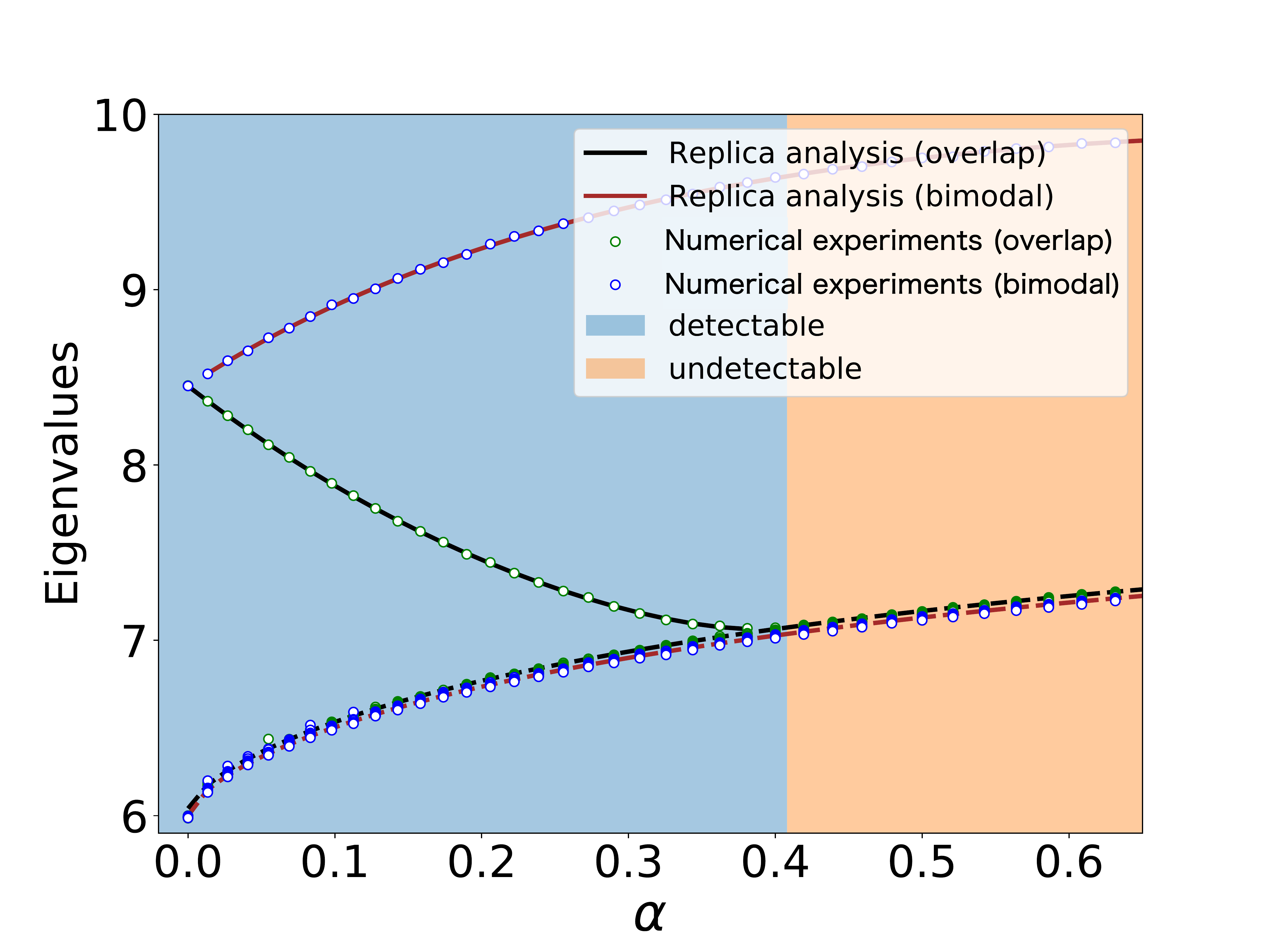}
\caption{Comparison between the overlapping and bimodal SBMs. This figure shows the eigenvalues of bimodal SBM in addition to those in Fig.~\ref{fig:top10_eigenvalues_overlap}. The blue dots represent the top ten eigenvalues of the bimodal SBM computed in the numerical experiments. The brown solid and dashed lines represent the leading eigenvalue and the bulk edge of the eigenvalues of the bimodal SBM that are derived by the replica analysis, respectively. The spectrum of the overlapping SBM is plotted as in  Fig.~\ref{fig:top10_eigenvalues_overlap}. These models are identical only when $\alpha=0$. However, their bulk edges should coincide when $\alpha=0$ and $\alpha=1$.
For the value of $\epsilon$ of the bimodal SBM, we used the same value as the overlapping SBM, which varies as $\alpha$ increases owing to constraint (\ref{eq:constraint_between_alp_eps}).}
\label{fig:top10_eigenvalues_overlap_vs_bimodal}
\end{figure}

We now investigate the effect of the overlapping structure on the accuracy of the spectral clustering when we increase the size of the overlapping block.
Because the overlapping block can have denser (or sparser) edge density than the other blocks, the average degree also increases (or decreases) accordingly, as the size of the overlapping block increases.
This implies that the width of the bulk of the eigenvalues is trivially influenced, because the bulk is known to depend on the average degree \cite{nadakuditi2012graph}.

However, it is not trivial if it is the only effect. Namely, the overlapping structure may affect the isolated eigenvalue or the bulk in another way.
To assess the effect of the overlapping structure rather than the effect of the average degree,  we compare the overlapping SBM with the model with no overlapping structures but that has the same degree distribution as the overlapping SBM. In the case of the microcanonical overlapping SBM, the degree distribution is bimodal: all the nodes in the overlapping block have the same degree, while all the other nodes have the other degree.
Therefore, we consider the non-overlapping SBM with a bimodal degree distribution (see Appendix~\ref{sec:appendix_bimodal} for a detailed definition). We assume that the sizes of the blocks are equal. Hereafter, we refer to this model as the bimodal SBM. 

Figure~\ref{fig:top10_eigenvalues_overlap_vs_bimodal} shows the bulks of eigenvalues and the leading eigenvalues of the overlapping and bimodal SBMs. We can confirm that both bulk edges almost coincide. In contrast, the leading eigenvalue of the bimodal SBM is separated from the bulk in the whole space, while that of the overlapping SBM approaches to its bulk as $\alpha$ increases. This indicates that the increase of the size of the overlapping block mainly affects the leading eigenvalue instead of the bulk.

The fact that the bulk is not considerably affected is not very trivial. If we take a closer look, the bulk edges do not exactly coincide in Fig.~\ref{fig:top10_eigenvalues_overlap_vs_bimodal}, although the deviation is very small. This is because the models are not identical even when there is no community structure (i.e., $\epsilon = 1$). 
When $\alpha=0$, the two models reduce to the $c_1$-regular SBM. Thereby, their bulk edges become equal to $2\sqrt{c_1-1}$. 
When $\alpha=1$, the overlapping SBM becomes a uniform (one block) model with (average) degree $c_2$, while the bimodal SBM has the community structure with (average) degree $c_2$. However, the bulk edge of the SBM with no overlapping blocks depends only on its average degree. Thus, although the models are not identical, their bulk edges are both $2\sqrt{c_2-1}$.

\subsection{Effects of the density of the overlapping structure}
Next, we investigate how density $\sigma$ of the overlapping block affects the detectability. As mentioned in the previous subsection, the higher density of the overlapping block trivially makes the width of the bulk of the eigenvalues expand wider.

\begin{figure*}[t!]
\centering
\begin{tabular}{c}

\begin{minipage}{0.45\hsize}
\includegraphics[clip, width=8.4cm]{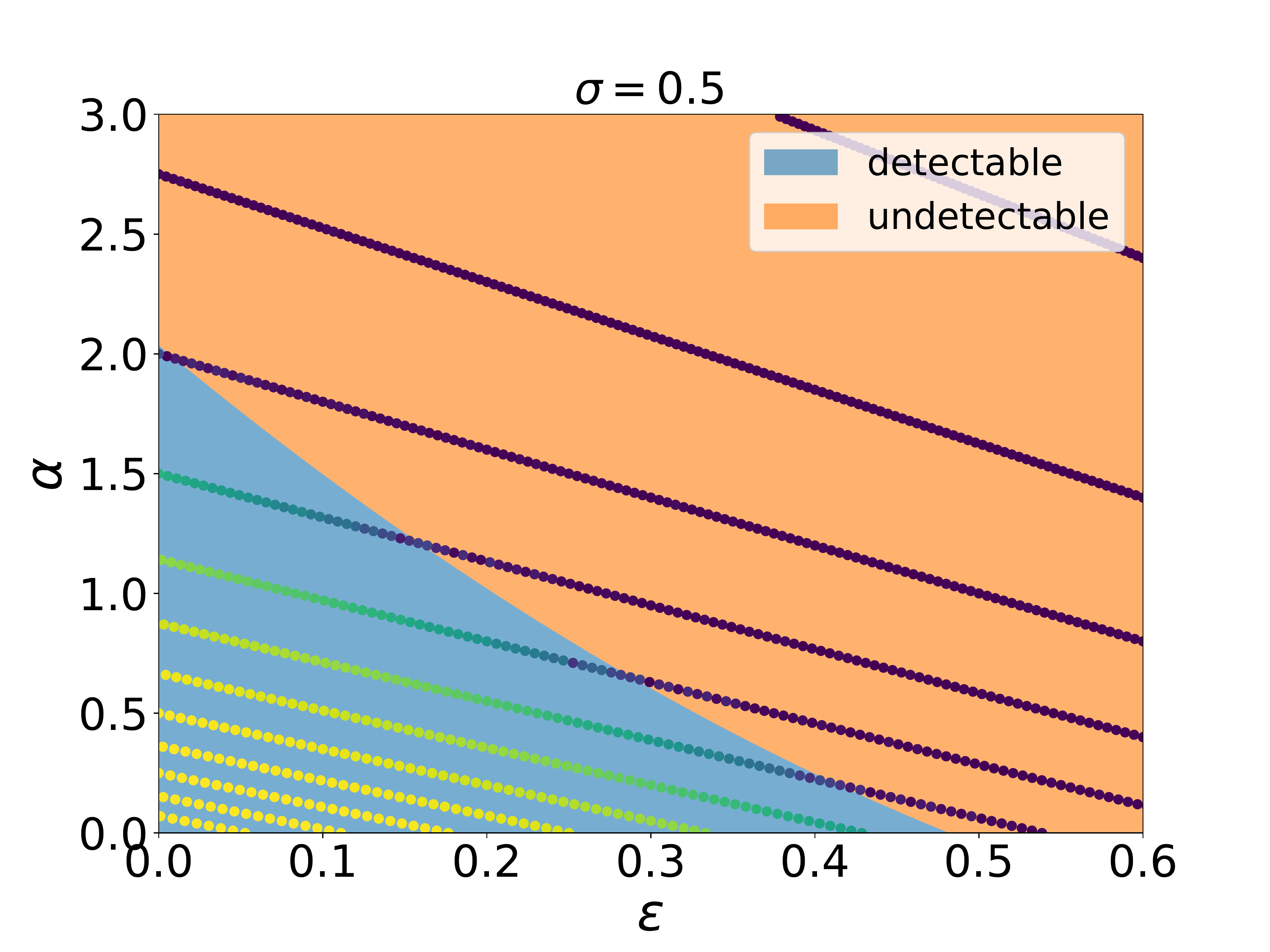}
\subcaption{}
\label{fig:phase_diagram_sigma_5}
\end{minipage}

\begin{minipage}{0.45\hsize}
\includegraphics[clip, width=8.4cm]{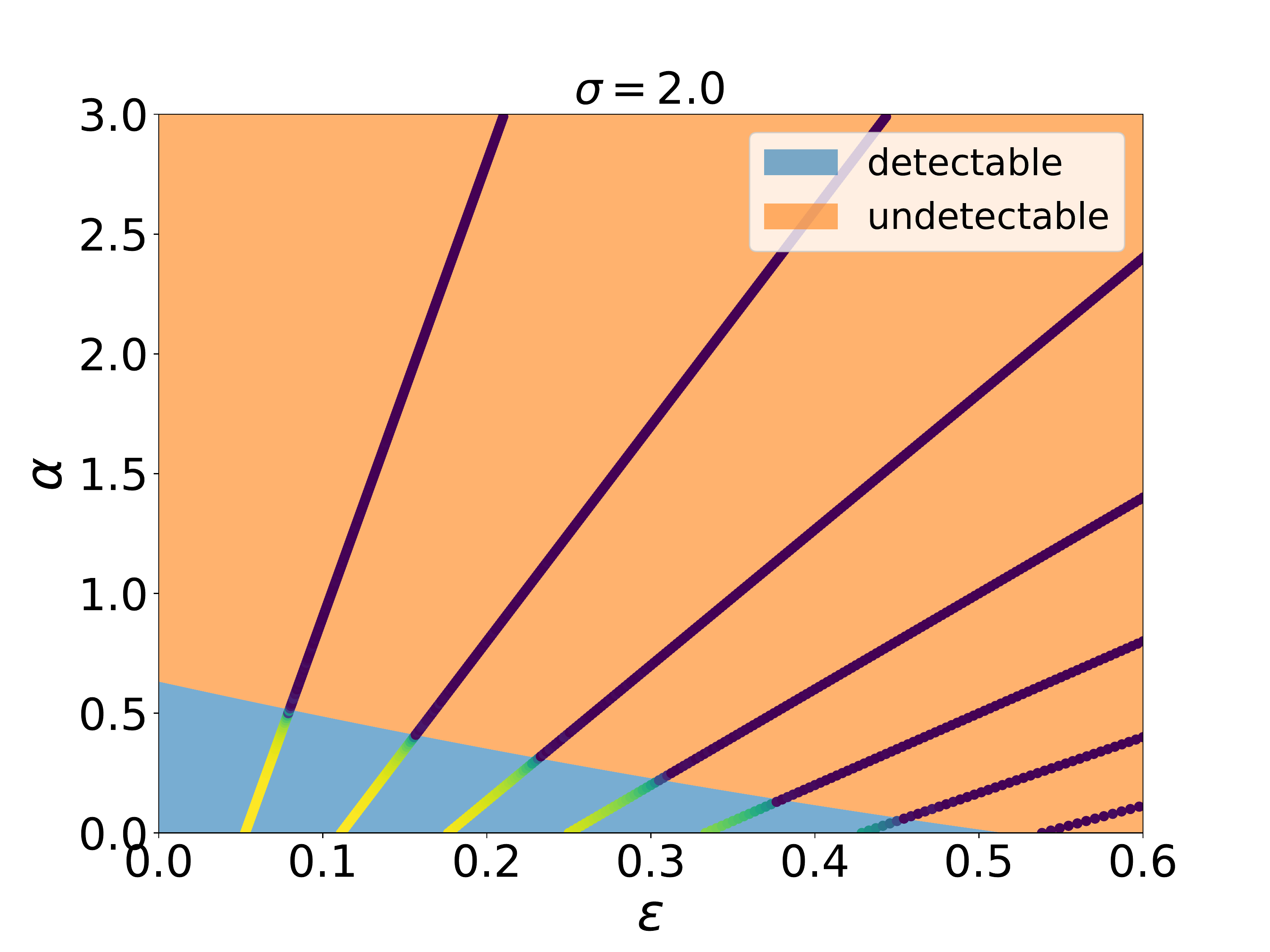}
\subcaption{}
\label{fig:phase_diagram_sigma20}
\end{minipage}

\end{tabular}
\caption{Detectability phase diagram of the $(\epsilon,\alpha)$ plane for $\sigma=0.5,\ 2$ and $c_1=10$. A detailed explanation is provided in the caption of Fig.~\ref{fig:phase_diagram}.}
\label{fig:phase_diagram_sigma_changed}
\end{figure*}

Figures~\ref{fig:phase_diagram_sigma_5}--\ref{fig:phase_diagram_sigma20} show the detectability phase diagram derived by the replica analysis and the results of the corresponding numerical experiments for $\sigma=0.5$ and $2$. 
Notably, the detectable region is wider when $\sigma$ is small. This indicates that the higher density deteriorates the detectability more significantly.

\begin{figure*}[t!]
\centering
\begin{tabular}{c}

\begin{minipage}{0.45\hsize}
\includegraphics[clip, width=8cm]{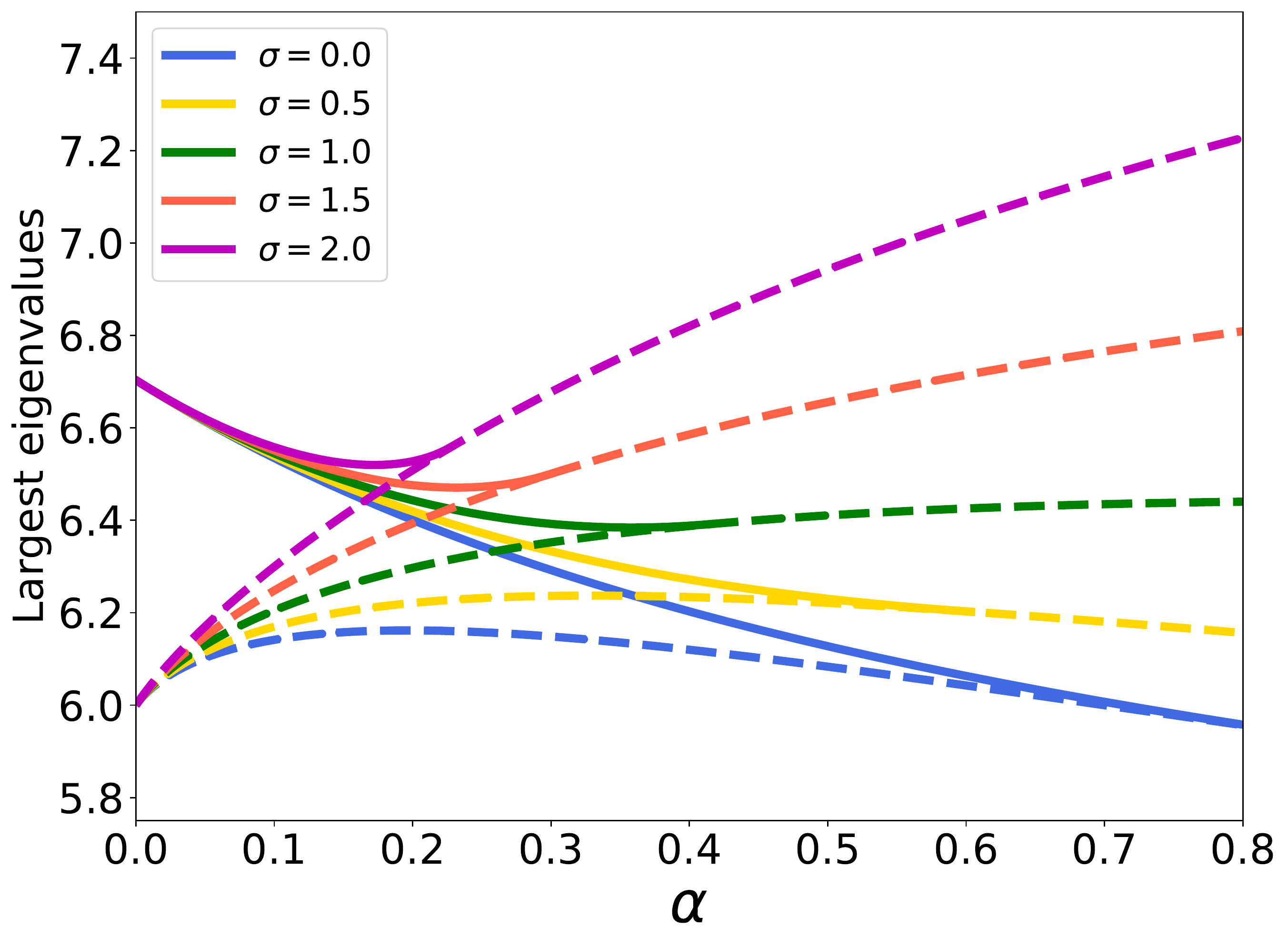}
\subcaption{}
\label{fig:largest_eigenvalues_sigma_changed}
\end{minipage}

\begin{minipage}{0.45\hsize}
\includegraphics[clip, width=8cm]{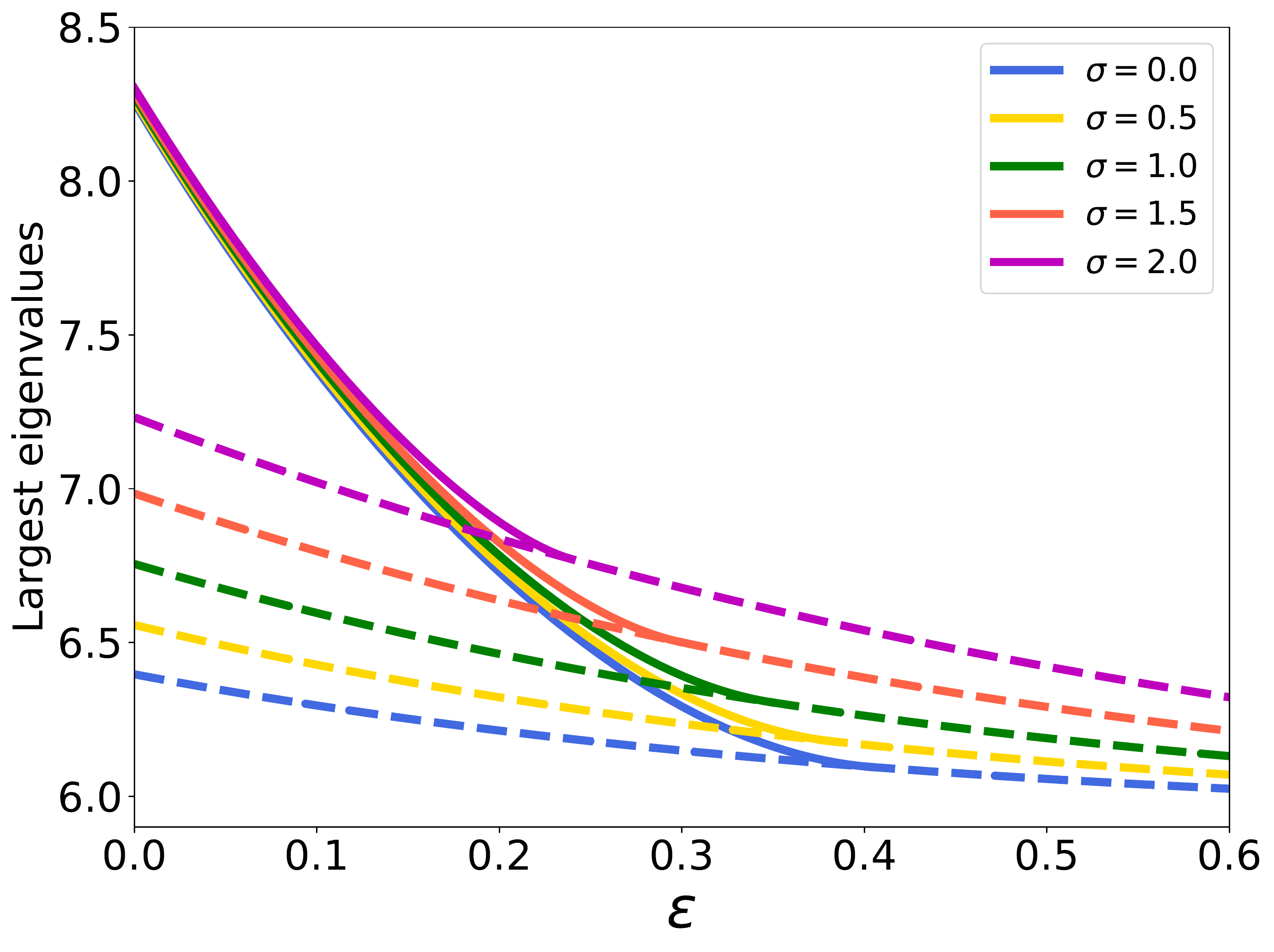}
\subcaption{}
\label{fig:largest_eigenvalues_delta_fixed_sigma_changed}
\end{minipage}

\end{tabular}
\caption{(a) Isolated eigenvalues (solid lines) and bulk edges (dashed lines) as a function of $\alpha$ for $\sigma=0, 0.5, 1, 1.5, 2$. Parameters are set to $c_1=10$ and $\epsilon=0.3$. The value of degree $c_2$ varies according to (\ref{eq:constraint_between_alp_eps}). (b) Isolated eigenvalues (solid lines) and bulk edges (dashed lines) as a function of $\epsilon$. $\alpha$ is fixed as 0.3. Other experimental conditions are identical to those of Fig.~\ref{fig:largest_eigenvalues_sigma_changed}.}
\label{fig:largest_eigenvalues_sigma_or_eps_changed}
\end{figure*}

Let us examine $\sigma$ dependency. 
Figure~\ref{fig:largest_eigenvalues_sigma_changed} shows the $\alpha$ dependencies derived by the replica analysis of the canonical SBM. They are the isolated leading largest eigenvalues and the bulk of the eigenvalues for $\sigma=0, 0.5, 1, 1.5$, and $2$. Interestingly, the isolated largest eigenvalue does not depend on $\sigma$ considerably. In contrast, the bulk is highly dependent on $\sigma$. This indicates that the deterioration of the detectability due to $\sigma$ is caused by the expansion of the bulk rather than the shrinkage of the isolated leading eigenvalue. Figure~\ref{fig:largest_eigenvalues_delta_fixed_sigma_changed} similarly shows the $\epsilon$ dependencies. Again, we can see that the isolated largest eigenvalue does not depend on $\sigma$ considerably while the bulk is highly dependent.

Notably, we cannot test the result of Fig.~\ref{fig:largest_eigenvalues_sigma_changed} directly in numerical experiments, because $\alpha$ cannot be varied continuously as $\epsilon$ is fixed. This is due to the constraints of the microcanonical SBM. Similarly, in Fig.~\ref{fig:largest_eigenvalues_delta_fixed_sigma_changed}, $\epsilon$ cannot be varied continuously as $\alpha$ is fixed. 
Nevertheless, we can draw smooth curves in the replica analysis, because we consider the canonical SBM that is not subject to the constraints of the microcanonical SBM. Importantly, the results of the microcanonical SBM coincide with those of the canonical SBM with the regular approximation at the points where the microcanonical SBM is realizable.
We also note that (Appendix~\ref{sec:appendix_derivation_micro}) the distinction between the canonical and microcanonical SBMs is invisible in infinite graph size limits.

\section{Summary}
\label{sec:summary}
We investigated the effect of the size and the density of the overlapping block on the accuracy of spectral clustering using the replica method. Both larger size and higher density help the isolated eigenvalue to be buried in the bulk of the eigenvalues, i.e., deteriorate the detectability. Importantly, however, their mechanisms are strikingly different. We found that increasing the size of the overlapping block has a prominent effect on making the isolated eigenvalue smaller (Fig.~\ref{fig:top10_eigenvalues_overlap_vs_bimodal}). In contrast, increasing the density of the overlapping block makes the bulk width larger, while the isolated eigenvalue remains almost the same (Fig.~\ref{fig:largest_eigenvalues_sigma_changed}). 

According to our findings, the results of the replica analysis are consistent with those of the numerical experiments. This indicates that the detectability phase transition of the spectral clustering in the present setting is regarded as a phenomenon that can be understood in the scope of the mean-field theory.

Although spectral clustering typically deals with non-overlapping structures, we showed that it is also possible to analyze the model-inconsistent case, such as the overlapping SBM. 
It is possible, in principle, to investigate even more complex situations using the replica method. However, for example, we would need to deal with saddle-point equations with many variables if we were to analyze a general three-block SBM. 
Therefore, we believe that the present model is an extreme case where the analytical calculation is executable and the results are interpretable. 

\section{Acknowledgements}
This study was funded by the New Energy and Industrial Technology Development Organization (NEDO), JSPS KAKENHI No. 18K18127 (T.K.) and JST CREST Grant Number JPMJCR1912.

\bibliography{mybib}
\appendix

\section{Derivation of the spectrum and the detectability limit of the canonical SBM}
\label{sec:appendix_b}
The goal of this appendix is to derive saddle-point expression of the average largest eigenvalue (\ref{eq:can_lam}). Note that a similar calculation using the replica method can be found in Refs.~\cite{kabashima2012first,kawamoto2015limitations,kawamoto2015detectability}. We start with the average of $n$th moment of the partition function
\begin{align}
    &[Z^n(M,\beta)]_M \nonumber \\
    &=
    \int \left(\prod_{a=1}^nd\boldsymbol{x}_a\delta(\boldsymbol{x}_a^\top\boldsymbol{x}_a-N)\right)
    \left[
    \exp\left(\frac{\beta}{2}\sum_a\boldsymbol{x}_a^\top M\boldsymbol{x}_a\right)\right]_M \label{eq:app_can_nth_moment_Z_mod}
    \\ &=
    \int \left(\prod_{a=1}^nd\boldsymbol{x}_a\delta(\boldsymbol{x}_a^\top\boldsymbol{x}_a-N)\right) \nonumber \\
    &\quad\times \left[\exp\left(-\frac{\beta}{2}\sum_a\left(\boldsymbol{\gamma}^\top\boldsymbol{x}_a\right)^2\right)
    \exp\left(\frac{\beta}{2}\sum_a\boldsymbol{x}_a^\top A\boldsymbol{x}_a\right)\right]_A,
    \label{eq:app_can_nth_moment_Z}
\end{align}
where $\gamma_i\equiv d_i/\sqrt{2m}$.
Introducing order parameters
\begin{equation}
    \Omega_a=\frac{1}{\sqrt{N}}\sum_i\gamma_ix_{ia},\ \ \  (a\in \{1,\dots,n\}),
    \label{eq:can_order_parameter_omega}
\end{equation}
we can recast exponential factor $e^{-\frac{\beta}{2}\sum_a(\boldsymbol{\gamma}^\top\boldsymbol{x}_a)^2}$ in (\ref{eq:app_can_nth_moment_Z}) as

\begin{align}
    &\exp\left(-\frac{\beta}{2}\sum_a\left(\boldsymbol{\gamma}^\top\boldsymbol{x}_a\right)^2\right)
    \nonumber\\
    &=\int \prod_ad\Omega_a\delta\left(\Omega_a-\frac{1}{\sqrt{N}}\sum_i\gamma_ix_{ia}\right)\exp{\left(-\frac{\beta N}{2}\sum_a\Omega_a^2\right)}\\
    &=\int \prod_a\frac{\beta N}{2\pi}d\Omega_ad\hat{\Omega}_ae^{-\frac{\beta N}{2}\sum_a(\Omega_a^2-2\Omega_a\hat{\Omega}_a)}e^{-\beta\sqrt{N}\sum_{ia}\hat{\Omega}_a\gamma_ix_{ia}}\\
    &=
    \int\prod_a\frac{\beta N}{2\pi}d\Omega_ad\hat{\Omega}_ae^{-\frac{\beta N}{2}\sum_a(\Omega_a^2-2\Omega_a\hat{\Omega}_a)}
    \prod_{ij}e^{-\frac{\beta}{\sqrt{\bar{c}}}\sum_a\hat{\Omega}_aA_{ij}x_{ia}}.
    \label{eq:can_exp_alp_x}
\end{align}
We have set $\bar{c}\equiv2m/N$. Moreover, $\hat{\Omega}_a$ is the auxiliary variable that is conjugate to $\Omega_a$. To derive this expression, we transformed the delta function to
\begin{align}
    &\delta\left(\sqrt{N}\Omega_a-\sum_i\gamma_ix_{ia}\right) \nonumber\\
    &=\int_{-i\infty}^{+i\infty}\frac{\beta\sqrt{N}}{2\pi}d\hat{\Omega}_ae^{\beta\sqrt{N}\hat{\Omega}_a(\sqrt{N}\Omega_a-\sum_i\gamma_ix_{ia})}.
    \label{eq:app_can_delta_function_omega}
\end{align}
Inserting Eq.~(\ref{eq:can_exp_alp_x}) into the exponential factor in (\ref{eq:app_can_nth_moment_Z}), we obtain
\begin{align}
    &\left[
    \exp\left(\frac{\beta}{2}\sum_a\boldsymbol{x}_a^\top M\boldsymbol{x}_a\right)\right]_M \nonumber\\
    &= \int\prod_a\frac{\beta N}{2\pi}d\Omega_a d\hat{\Omega}_ae^{-\frac{\beta N}{2}(\Omega_a^2-2\Omega_a\hat{\Omega}_a)} \nonumber\\
    &\quad\times\prod_{i<j}\sum_{A_{ij}\in\{0,1\}}\rho_{t_it_j}^{A_{ij}}(1-\rho_{t_it_j})^{1-A_{ij}}
    e^{\beta\sum_aA_{ij}x_{ia}\left(x_{ja}-\frac{2\hat{\Omega}_a}{\sqrt{\bar{c}}}\right)} \label{eq:can_after_alp_break1}\\
    &\approx
    \int\prod_a\frac{\beta N}{2\pi}d\Omega_a d\hat{\Omega}_ae^{-\frac{\beta N}{2}(\Omega_a^2-2\Omega_a\hat{\Omega}_a)} \nonumber\\
    &\quad\times
    \prod_{i<j}
    \exp\left(\log(1-\rho_{t_it_j})+\rho_{t_it_j}e^{\beta\sum_ax_{ia}(x_{ja}-\frac{2\hat{\Omega}_a}{\sqrt{\bar{c}}})}\right).
    \label{eq:can_after_alp_break2}
\end{align}
Here, we took the configuration average over the canonical SBM (\ref{eq:canonical_sbm_probability}) and approximated $\frac{\rho_{t_it_j}}{1-\rho_{t_it_j}}\approx\rho_{t_it_j}$ by using the fact that $\rho_{t_it_j}=O(N^{-1})$.

Let us now introduce the order-parameter functions
\begin{equation}
    Q_k(\boldsymbol{u})=\frac{1}{p_kN}\sum_{i\in V_k}\prod_a\delta(u_a-x_{ia}),\ \ \  \left(k\in\{1,\dots,K\}\right)
    \label{eq:app_order_parameter_function}
\end{equation}
where $\sum_{i\in V_k}$ is the sum over the node indices that belong to the $k$th block.
Then, the last exponential factor in (\ref{eq:can_after_alp_break2}) can be approximated as
\begin{align}
    &\exp\left(\sum_{i<j}\rho_{t_it_j}e^{\beta\sum_ax_{ia}(x_{ja}-\frac{2\hat{\Omega}_a}{\sqrt{\bar{c}}})}\right) \nonumber\\
    &\approx
    \exp\left(
    \frac{N^2}{2}\int\prod_adu_adv_ae^{\beta u_a\left(v_a-\frac{2\hat{\Omega}_a}{\sqrt{\bar{c}}}\right)}\right.\nonumber \\
    &\qquad\qquad\left.\times\sum_{kk'}Q_k(\boldsymbol{u})W_{kk'}Q_{k'}(\boldsymbol{u})
    \right),
    \label{eq:can_after_order_parameter_fucntion}
\end{align}
where we approximated that the contribution from the diagonal elements is negligible, and we defined $W_{kk'}\equiv p_k\rho_{kk'}p_{k'}$.
Inserting Eq.~(\ref{eq:can_after_order_parameter_fucntion}) into (\ref{eq:can_after_alp_break2}), Eq.~(\ref{eq:app_can_nth_moment_Z_mod}) is now expressed as
\begin{align}
    &[Z^n(M,\beta)]_M \nonumber\\
    &=
    \int\prod_ad\boldsymbol{x}_a\int\prod_a\frac{\beta N}{2\pi}d\hat{\Omega}_ad\Omega_a
    \prod_a\delta(\boldsymbol{x}_a\boldsymbol{x}_a^{\top}-N)
    \nonumber\\
    &\times
    \exp
    \left(
    -\frac{\beta N}{2}\sum_a(\Omega_a^2-2\Omega_a\hat{\Omega}_a)
    +\sum_{i<j}\log(1-\rho_{t_it_j})
    \right.\nonumber \\
    &+\left.
    \frac{N^2}{2}\int\prod_adu_adv_ae^{\beta u_a\left(v_a-\frac{2\hat{\Omega}_a}{\sqrt{\bar{c}}}\right)}
    \sum_{kk'}Q_k(\boldsymbol{u})W_{kk'}Q_{k'}(\boldsymbol{u})
    \right).
    \label{eq:app_can_nth_moment_before_transformation}
\end{align}
Here, we use the expansion of the delta function
\begin{equation}
    \delta\left(\boldsymbol{x}_a^{\top}\boldsymbol{x}_a-N\right)
    =\int_{-i\infty}^{+i\infty}\frac{\beta d\phi_a}{4\pi}e^{-\frac{\beta}{2}\phi_a\left(\sum_ix_{ia}^2-N\right)}
    \label{eq:app_can_delta_function_phi}
\end{equation}
and the identity
\begin{align}
    1&=\prod_kp_kN\int\frac{D{Q}_k}{2\pi} \nonumber\\
    &\quad\times\delta\left(\sum_{i\in V_k}z_i\prod_{a=1}^n\delta(x_{ia}-\mu_a)-p_kN{Q}_k(\boldsymbol{\mu})\right)\label{eq:can_order_parameter_function_identity} \\
    &=
    \prod_kp_kN\int\frac{D{Q}_kD\hat{{Q}}_k}{2\pi}\exp\left(\sum_k\int d\boldsymbol{\mu}\hat{{Q}}_k(\boldsymbol{\mu})\right.\nonumber \\
    & \quad \left.\times\left(\sum_{i\in V_k}z_i\prod_{a=1}^n\delta(x_{ia}-\mu_a)-p_kN{Q}_k(\boldsymbol{\mu})\right)\right).
    \label{eq:app_can_order_parameter_function_identity2}
\end{align}
Here, $\int DQ_k$ is the functional integral with respect to ${Q}_k(\boldsymbol{\mu})$, and $\hat{{Q}}_k(\boldsymbol{\mu})$ was introduced as the conjugate of ${Q}_k(\boldsymbol{\mu})$.
To derive Eq.~(\ref{eq:app_can_order_parameter_function_identity2}), we used the expansion of the delta function.
By inserting the identity, we can focus on ${Q}_k(\boldsymbol{\mu})$ corresponding to the replacement in (\ref{eq:app_order_parameter_function}). Note that without the insertion of the identity, the replacement of (\ref{eq:app_order_parameter_function}) becomes invalid. From these, we can recast Eq.~(\ref{eq:app_can_nth_moment_before_transformation}) as 
\begin{align}
    & [Z^n(M,\beta)]_M \nonumber\\
    &=
    \int\prod_a\frac{d\phi_a}{4\pi}\int\prod_a\frac{\beta N}{2\pi}d\hat{\Omega}_ad\Omega_a\int\prod_k\frac{p_kN}{2\pi}D\hat{Q}_kDQ_k\nonumber \\
    &\times\exp
    \left(
    -\frac{\beta N}{2}\sum_a(\Omega_a^2-2\Omega_a\hat{\Omega}_a-\phi_a)
    +\sum_{i<j}\log(1-\rho_{t_it_j})\right.\nonumber \\
    &\quad-\left.\sum_kp_kNL_{k}(Q_k,\hat{Q}_k)+K(\{Q_k\})\right.\nonumber \\
    &\quad\left.+\sum_k\sum_{i\in V_k}\log M_{i,k}(\hat{Q}_k,\{\phi_a\})
    \right),
    \label{eq:can_K_nth_after_order_parameter_function}
\end{align}
where 
\begin{align}
    K(\{Q_k\})&=\frac{N^2}{2}\int\prod_adu_adv_ae^{\beta u_a\left(v_a-\frac{2\hat{\Omega}_a}{\sqrt{\bar{c}}}\right)} \nonumber\\
    &\quad\times\sum_{kk'}Q_k(\boldsymbol{u})W_{kk'}Q_{k'}(\boldsymbol{v}),\label{eq:can_K} \\
    L_k(Q_k,\hat{Q}_k)&=\int d\boldsymbol{u}Q_k(\boldsymbol{u})\hat{Q}_k(\boldsymbol{u}),\label{eq:can_Lk} \\
    M_{i,k}(\hat{Q}_k,\{\phi_a\})&=\int\prod_ad\boldsymbol{x}_ae^{\hat{Q}_k(\boldsymbol{x}_i)-\frac{\beta}{2}\sum_a\phi_ax_{ia}^2}. \label{eq:can_Mik}
\end{align}

Here, we assume the functional form of ${Q}_k(\boldsymbol{u})$ and $\hat{Q}_k(\boldsymbol{u})$ are restricted to Gaussian mixtures. This indicates that ${Q}_k(\boldsymbol{u})$ and $\hat{Q}_k(\boldsymbol{u})$ can be expressed as
\begin{align}
    {Q}_k(\boldsymbol{u})&=q_k^0\int dAdHq_k(A,H)\left(\frac{\beta A}{2\pi}\right)^{\frac{n}{2}} \nonumber\\
    &\quad\times\exp\left(-\frac{\beta A}{2}\sum_a\left(\mu_a-\frac{H}{A}\right)^2\right), \label{eq:can_gaussian_mix} \\
    \hat{{Q}}_k(\boldsymbol{u})&=\hat{q}_k^0\int d\hat{A}d \hat{H}\hat{q}_k(\hat{A}, \hat{H})\exp \left(\frac{\beta}{2}\sum_a\left(\hat{A}\mu_a^2+2\hat{H}\mu_a\right)\right),
    \label{eq:can_gaussian_mix_hat}
\end{align}
where $q_k(A,H)$ is the weight of a Gaussian distribution with the mean and precision parameter equal to $H/A$ and $H$, respectively. $\hat{q}_k(\hat{A}, \hat{H})$ is defined analogously. $q_k^0$ and $\hat{q}_k^0$  are the normalization constants; it can be deduced that $q_k^0=1$ and $\hat{q}_k^0=c_k$ from the saddle-point conditions when $n=0$.
Inserting Eq.~(\ref{eq:can_gaussian_mix}) and (\ref{eq:can_gaussian_mix_hat}) into (\ref{eq:can_K})--(\ref{eq:can_Mik}), we have
\begin{widetext}
\begin{align}
    K(\{Q_k\})&=\frac{N^2}{2}\sum_{kk'}W_{kk'}\int dAdHq_k(A,H)\int dA'dH'q_{k'}(A',H')\left(\frac{AA'}{AA'-1}\right)^{-\frac{n}{2}} \nonumber \\
    &\times \exp\left[
    \sum_a\frac{\beta}{2}
    \left(
    \frac{A'\left(H-\frac{2\hat{\Omega}_a}{\sqrt{\bar{c}}}\right)^2+2\left(H-\frac{2\hat{\Omega}}{\sqrt{\bar{c}}}\right)H'+AH'^2}{AA'-1}-\frac{\left(H-\frac{2\hat{\Omega}_a}{\sqrt{\bar{c}}}\right)^2}{A}-\frac{H'^2}{A}
    \right)
    \right], \label{eq:can_K_after_gm}
\end{align}

\begin{equation}
    L_k(Q_k,\hat{Q}_k)=\int dAdHd\hat{A}\hat{H}
    q_k(A,H)\hat{q}_k(\hat{A},\hat{H})\left(\frac{A}{A-\hat{A}}\right)^{\frac{n}{2}}\exp\left[\frac{n\beta}{2}\left(\frac{(H+\hat{H})^2}{A-\hat{A}}-\frac{H^2}{A}\right)\right],
\end{equation}
\begin{equation}
    M_{i,k}(\hat{Q}_k,\{\phi_a\}) =\left(\frac{2\pi}{\beta}\right)^{\frac{n}{2}}\sum_{d=0}^\infty\frac{c_k^d}{d!}\int \prod_{g=1}^d\left(d\hat{A}_gd\hat{H}_g\hat{q}_k(\hat{A}_g,\hat{H}_g)\right)
    \prod_a\left(\phi_a-\sum_g\hat{A}_g\right)^{-\frac{1}{2}}
    \exp\left(\frac{\beta}{2}\frac{\left(\sum_g\hat{H}_g\right)^2}{\phi_a-\sum_g\hat{A}_g}\right).
    \label{eq:can_Mik_after_gm}
\end{equation}
\end{widetext}
To derive Eq.~(\ref{eq:can_Mik_after_gm}), we expanded the exponential as $e^{\hat{Q}_k(\boldsymbol{x}_i)}=\sum_{d=0}^\infty\frac{1}{d!}\hat{Q}_k^d(\boldsymbol{x}_i)$.

Hereafter, let us assume no distinction among the variables with different replica indices, i.e., $\phi_a=\phi$, $\Omega_a=\Omega$, and $\hat{\Omega}_a=\hat{\Omega}$. This is referred to as the {\it replica symmetric assumption}. We insert Eq.~(\ref{eq:can_K_after_gm})--(\ref{eq:can_Mik_after_gm}) into (\ref{eq:can_K_nth_after_order_parameter_function}) under this assumption. Then, we obtain the following saddle-point equation for the average largest eigenvalue from Eqs.~(\ref{eq:free_energy_log_z}), (\ref{eq:beta_inf_free_energy}), and (\ref{eq:replica_trick}) as
\begin{widetext}
\begin{align}
    [\lambda(M)]_M 
    &= 2\lim_{\beta\to\infty}\frac{1}{\beta N}\lim_{n\to0}\frac{\partial}{\partial n}\log [Z^n]_M \\
    &= \underset{\phi,\Omega,\hat{\Omega},\{q_k,\hat{q}_k\}}{\operatorname{extr}}\left[
    \phi+2\Omega\hat{\Omega}-\Omega^2
    +\frac{1}{2}\sum_{kk'}NW_{kk'}\int dAdHq_k(A,H)\int dA'dH'q_{k'}(A',H')\right.\nonumber \\
    &\quad\times\left(
    \frac{A'\left(H-\frac{2\hat{\Omega}}{\sqrt{\bar{c}}}\right)^2+2\left(H-\frac{2\hat{\Omega}}{\sqrt{\bar{c}}}\right)H'+AH'^2}{AA'-1}-\frac{(H-\frac{2\hat{\Omega}}{\sqrt{\bar{c}}})^2}{A}-\frac{H'^2}{A'}
    \right)\nonumber \\
    & -\sum_kp_kc_k\int dAdHd\hat{A}d\hat{H}q_k(A,H)\hat{q}_k(\hat{A},\hat{H})\left(\frac{(H+\hat{H})^2}{A-\hat{A}}-\frac{H^2}{A}\right)\nonumber \\
    &+\left.\sum_kp_k\sum_{d=0}^\infty\mathcal{P}_{c_k}(d)
    \int\prod_{g=1}^d(d\hat{A}_gd\hat{H}_g\hat{q}_k(\hat{A}_g,\hat{H}_g))\frac{\left(\sum_g\hat{H}_g\right)^2}{\phi-\sum_g\hat{A}_g}\right].
    \label{eq:can_lam_before_ema}
\end{align}
\end{widetext}
Here, $\mathcal{P}_{c_k}(d)$ is the probability mass function of degree $d$ of each node in block $k$ that has expectation $c_k$.
From the saddle-point condition in Eq.~(\ref{eq:can_lam_before_ema}), we obtain the functional equations with respect to $q_k(A,H)$ and $\hat{q}_k(\hat{A},\hat{H})$ as
\begin{align}
    q_k(A,H)&=\sum_{d=0}^\infty\mathcal{P}_{c_k}(d)d\int\prod_{g=1}^{d-1}
    \left(d\hat{A}_gd\hat{H}_g\hat{q}_k(\hat{A}_g,\hat{H}_g)\right) \nonumber\\
    &\quad\times\delta\left(H-\sum_{g=1}^{d-1}\hat{H}_g\right)
    \delta\left(A-\phi+\sum_{g=1}^{d-1}\hat{A}_g\right), \label{eq:can_func_saddle_equation1}\\
    \hat{q}_k(\hat{A},\hat{H})&=\frac{1}{c_k}\sum_{k'}N\rho_{kk'}p_{k'}
    \int dA'dH'q_{k'}(A',H') \nonumber\\
    & \quad\times\delta\left(\hat{A}-\frac{1}{A'}\right)
    \delta\left(\hat{H}-\frac{H'-\frac{2\hat{\Omega}}{\sqrt{\bar{c}}}}{A'}\right).
    \label{eq:can_func_saddle_equation2}
\end{align}
To derive Eq.~(\ref{eq:can_func_saddle_equation1}), we used the fact that the expectation of $H^2/A$ becomes 0, which is derived by substituting $\hat{H}=\hat{A}=0$.
Moreover, the saddle-point condition with respect to $\phi$ yields
\begin{equation}
    \sum_kp_k\int dAdH\mathcal{Q}_k(A,H)\left(\frac{H}{A}\right)^2=1,
    \label{eq:can_func_constraint}
\end{equation}
where 
\begin{align}
    \mathcal{Q}_k(A,H)&=\sum_{d=0}^\infty\mathcal{P}_{c_k}(d)\int\prod_{g=1}^d
    \left(d\hat{A}_gd\hat{H}_{g}\hat{q}_k(\hat{A}_g,\hat{H}_g)\right) \nonumber\\
    &\quad\times\delta\left(H-\sum_{g=1}^d\hat{H}_g\right)
    \delta\left(A-\phi+\sum_{g=1}^d\hat{A}_g\right).
\end{align}
Equation~(\ref{eq:can_func_constraint}) corresponds to the normalization constraint in (\ref{eq:modularity_maximization_problem2}). Equations~(\ref{eq:can_func_saddle_equation1}) and (\ref{eq:can_func_saddle_equation2}) constitute functional equations under constraint (\ref{eq:can_func_constraint}), and solving these equations yields the distribution of the largest eigenvector elements. Note that $q_k(A,H)$ was introduced as the weight in the Gaussian mixture, which approximates the empirical distribution of the largest eigenvector elements in (\ref{eq:app_order_parameter_function}). This indicates that $q_k(A,H)$ exhibits the probability density of the eigenvector-element distribution.

Unfortunately, solving the functional form of equations is still not analytically tractable. Thus, we introduce further approximations that $q_k(A)=\delta(A-a_k)$ and $\hat{q}_k(\hat{A})=\delta(\hat{A}-\hat{a}_k)$, i.e., we ignore the fluctuation of the precision parameters. This is called the {\it effective medium approximation} (EMA) \cite{biroli1999single,kabashima2012first}. Performing the EMA for (\ref{eq:can_lam_before_ema}), we arrive at
\begin{widetext}
\begin{align}
    \left[ \lambda(M) \right] _ { M } 
    &= \underset{\phi, \Omega, \hat{\Omega}, m_{1k}, m_{2k},\hat{m}_{1k}, \hat{m}_{2k}, a_k,\hat{a}_k}{\operatorname { extr }} \left\{ \phi + 2 \hat { \Omega } \Omega - \Omega ^ { 2 }\right. \nonumber \nonumber\\
   &\ + \frac { 1 } { 2 }N \sum _ { k , k ^ { \prime } } W _ { k k ^ { \prime } } \left( \frac { a_{k'} 
   \left(m _ { 2 k }-\frac{2\hat{\Omega}}{\sqrt{\bar{c}}}+\frac{4\hat{\Omega}^2}{\bar{c}}\right)
   + 2 m _ { 1 k' }
   \left(m_{1k}-\frac{2\hat{\Omega}}{\sqrt{\bar{c}}}\right)
   + a_{k} m _ { 2 k ^ { \prime } } } { a_k a_{k'} - 1 }
   -\frac{m_{2k}-\frac{2\hat{\Omega}}{\sqrt{\bar{c}}}m_{1k}+\frac{4\hat{\Omega}}{\bar{c}}}{a_k}-\frac{m_{2k'}}{a_{k'}}\right) \nonumber\\
    &\ - \sum _ { k } p _ { k } c _ { k } \left( \frac { m _ { 2 k } + 2 m _ { 1 k } \hat { m } _ { 1 k } + \hat { m } _ { 2 k } } { a_k - \hat { a }_k } -\frac{m_{2k'}}{a_{k'}} \right) + \left.\frac { 1 } { N } \sum _ { k } \sum _ { i \in V _ { k } } \sum _ { d = 0 } ^ { \infty } \frac { \mathcal { P } _ { c _ { k } } ( d ) } { \phi - d \hat { a }_k } \left( d \hat { m } _ { 2 k } + d ( d - 1 ) \hat { m } _ { 1 k } ^ { 2 } \right) \right\},
    \label{eq:app_can_lam}
\end{align}
\end{widetext}
where $m_{\ell k}$ and $\hat{m}_{\ell k}$ stand for the $\ell$th moments of $H$ and $\hat{H}$, respectively, i.e., $m_{\ell k}=\int dHH^\ell q_k(H)$ and $\hat{m}_{\ell k}=\int d\hat{H}\hat{H}^\ell\hat{q}_k(\hat{H})$. 

The saddle-point conditions from (\ref{eq:app_can_lam}) lead to the equations for the auxiliary variables $\phi, \Omega, \hat{\Omega},m_{\ell k},\hat{m}_{\ell k},a_k$, and $\hat{a}_k$. Here, we focus on a model with the symmetry between the community blocks: $p_1=p_3$ and $c_1=c_3$. Due to this assumption, we can apply the same assumptions to the physical quantities $a_k, \hat{a}_k, m_{2k}, \hat{m}_{2k}$, that is, $a_1=a_3$, $\hat{a}_1=\hat{a}_3$, $m_{21}=m_{23}$, and $\hat{m}_{21}=\hat{m}_{23}$. This is because these quantities are the second-order statistics and do not depend on the signs. 

Further, we assume $m_{12}=0$. This assumption stems from the fact that the overlapping block does not contain nodes in the community blocks. Thus, the corresponding elements of the eigenvector come from a random structure of the graph. Moreover, we classify the solution into the cases of $m_{11}=0$ and $m_{11}\neq 0$. For the solution with $m_{11}=0$, we can assume $m_{13}=0$ owing to the symmetry. On the other hand, for the solutions with $m_{11}\neq 0$, we can assume $m_{11}=-m_{13}$ due to the symmetry and the fact that the eigenvector elements of $\boldsymbol{x}$ tend to have the same signs in the same block. In summary, we have two types of solutions: $m_{11}=-m_{13}\neq 0,\ m_{12}=0$ and $m_{11}=m_{12}=m_{13}=0$.
In fact, the former corresponds to the detectable condition and the latter corresponds to the undetectable condition. The leading eigenvalue is calculated for each of the two conditions, and the detectability limit is derived as the boundary between these two conditions. We further simplify the problem using the regular approximation with respect to the degree, namely the random variables following the Poisson distribution $d$ in (\ref{eq:can_lam}) are fixed as their means $c_k$.

First, under the detectable condition, we can derive the equations for $a_1,a_2,\hat{a}_1$, and $\hat{a}_2$ from the saddle-point conditions as
\begin{align}
    a_1 + (c_1-1)\hat{a}_1 &= a_2 + (c_2-1)\hat{a}_2,\label{eq:can_a_ah_eq_1} \\
    \frac{1}{a_1-\hat{a}_1} &= \frac{1+\epsilon}{1+\alpha+\epsilon}\frac{a_1}{a_1^2-1}+\frac{\alpha}{1+\alpha+\epsilon}\frac{a_2}{a_1a_2-1}, \\
    \frac{1}{a_2-\hat{a}_2} &= \frac{\sigma\alpha}{\sigma\alpha+2}\frac{a_2}{a_2^2-1}+\frac{2}{\sigma\alpha+2}\frac{a_1}{a_1a_2-1}. \label{eq:can_a_ah_eq_3} \\
    \frac{1}{a_1-\hat{a}_1} &= \frac{1-\epsilon}{1+\epsilon+\alpha}\frac{c_1-1}{a_1^2-1}. \label{eq:can_a_ah_eq_4}
\end{align}
We let the solutions of Eq.~(\ref{eq:can_a_ah_eq_1})--(\ref{eq:can_a_ah_eq_4}) as $a_1^{\operatorname{det}}$, $a_2^{\operatorname{det}}$, $\hat{a}_1^{\operatorname{det}}$, and $\hat{a}_2^{\operatorname{det}}$. Then, we obtain the average leading eigenvalue as
\begin{equation}
    [\lambda(M)]_M=\phi=a_k^{\operatorname{det}}+(c_k-1)\hat{a}_k^{\operatorname{det}}\ \ \ (k=1,2)
    \label{eq:can_larget_eigenvalue_det}
\end{equation}
and the condition of the detectability limit as
\begin{equation}
    D(a_1^{\operatorname{det}},a_2^{\operatorname{det}},\hat{a}_1^{\operatorname{det}},\hat{a}_2^{\operatorname{det}})=0,
    \label{eq:can_detectability_threshold}
\end{equation}
where
\begin{equation}
    D(a_1,a_2,\hat{a}_1,\hat{a}_2)=M_{11}M_{22} - M_{12}M_{21},
\end{equation}
\begin{align}
    M_{11}&= (1+\epsilon)\frac{a_1^2+1}{(a_1^2-1)^2}+\alpha\frac{a_2^2}{(a_1a_2-1)^2} \nonumber\\
    &\quad-(1+\alpha+\epsilon)\frac{1}{(a_1-\hat{a}_1)^2}\frac{c_1}{c_1-1}, \label{eq:can_M11}\\
    M_{12} &=\frac{\alpha}{(a_1a_2-1)^2}, \label{eq:can_M12} \\
    M_{21} &=\frac{2}{(a_1a_2-1)^2},\label{eq:can_M21} \\
    M_{22} &= \frac{2a_1^2}{(a_1a_2-1)^2}+\sigma\alpha\frac{a_2^2+1}{(a_2^2-1)^2} \nonumber\\
    &\qquad-(\sigma\alpha+2)\frac{1}{(a_2-\hat{a}_2)^2}\frac{c_2}{c_2-1}. \label{eq:can_M22}
\end{align}
The detectability limit (\ref{eq:can_detectability_threshold}) is derived by condition $\hat{m}_{11}^2= 0$, because $D(a_1,a_2,\hat{a}_1,\hat{a}_2)$ is proportional to $\hat{m}_{11}^2$.

Second, under the undetectable condition, we can derive the equations for $a_1, a_2, \hat{a}_1$, and $\hat{a}_2$ from the saddle-point conditions as
\begin{align}
    a_1 + (c_1-1)\hat{a}_1 &= a_2 + (c_2-1)\hat{a}_2,\label{eq:can_a_ah_eq_1_und} \\
    \frac{1}{a_1-\hat{a}_1} &= \frac{1+\epsilon}{1+\alpha+\epsilon}\frac{a_1}{a_1^2-1}+\frac{\alpha}{1+\alpha+\epsilon}\frac{a_2}{a_1a_2-1}, \\
    \frac{1}{a_2-\hat{a}_2} &= \frac{\sigma\alpha}{\sigma\alpha+2}\frac{a_2}{a_2^2-1}+\frac{2}{\sigma\alpha+2}\frac{a_1}{a_1a_2-1}, \label{eq:can_a_ah_eq_3_und} \\
    D(a_1,a_2,\hat{a}_1,\hat{a}_2)&=0. \label{eq:can_a_ah_eq_4_und}
\end{align}
These equations are analogous to those for the detectable conditions (\ref{eq:can_a_ah_eq_1})--(\ref{eq:can_a_ah_eq_4}). A crucial difference is that we have condition $\hat{m}_{11}^2=0$ instead of Eq.~(\ref{eq:can_a_ah_eq_4}). We let the solutions of these equations be $a_1^{\operatorname{und}}$, $a_2^{\operatorname{und}}$, $\hat{a}_1^{\operatorname{und}}$, and $\hat{a}_2^{\operatorname{und}}$. Using this solution, we obtain the average leading eigenvalue in the undetectable conditions as follows.
\begin{equation}
    [\lambda(M)]_M=\phi=a_k^{\operatorname{und}}+(c_k-1)\hat{a}_k^{\operatorname{und}}.\ \ \ (k=1,2)
    \label{eq:can_largest_eigenvalue_und}
\end{equation}

\section{Microcanonical overlapping SBM}
\label{sec:micro_stochas_block_model}
In this appendix, we discuss the microcanonical SBM. In Sec.~\ref{sec:appendix_micro_model_definition}, we introduce the definition of the microcanonical overlapping SBM. In Sec.~\ref{sec:appendix_derivation_micro}, we provide the replica analysis to derive its spectrum and the detectability limit. In Sec.~\ref{sec:appendix_A}, we derive the saddle-point conditions for normalization constant $\mathcal{N}_G$, from which we can derive crucial relations used in Sec.~\ref{sec:appendix_derivation_micro}.
Finally, in Sec.~\ref{sec:appendix_comparison_two_models}, we discuss the distinction between the canonical and microcanonical SBMs and discuss the reason of their use in our numerical experiments.

\subsection{Model definition}
\label{sec:appendix_micro_model_definition}
Microcanonical SBM is an SBM that is formulated on the basis of different constraints from its canonical model.
Although the canonical SBM specifies the expected number of edges within the blocks, the microcanonical SBM specifies the number of edges within the blocks as well as the degree sequence as hard constraints. The microcanonical SBM generates a graph uniformly and randomly from all realizable graphs under these constraints.
We denote the sequence of node degrees as $\boldsymbol{d}=[d_i]$. We let $e_{kl}$ be the number of edges between blocks $k$ and $l$; we denote the corresponding matrix as $\boldsymbol{e}=[e_{kl}]$. Moreover, 
$\boldsymbol{t}=[t_i]$ $t_i\in\{1,\cdots,K\}$ $(i\in V)$ are the planted block labels of the nodes.
An instance of the microcanonical SBM is generated according to the following probability distribution.
\begin{equation}
    P(A|\boldsymbol{d},\boldsymbol{e},\boldsymbol{t})=\frac{1}{\Omega(\boldsymbol{d},\boldsymbol{e},\boldsymbol{t})},
    \label{eq:micro_distribution}
\end{equation}
where $\Omega(\boldsymbol{d},\boldsymbol{e},\boldsymbol{t})$ is the number of all realizable graphs under given $\boldsymbol{d}$, $\boldsymbol{e}$, and $\boldsymbol{t}$.

We consider a microcanonical SBM with an overlapping structure with the following parametrization.
\begin{align}
    \boldsymbol{p}&=(p_1,p_2,p_3)=\left(p_1,\alpha p_1,p_1\right),\label{eq:micro_group_sizes} \\
    \boldsymbol{e} &=
    \begin{pmatrix}
        1 & \alpha & \epsilon \\
        \alpha & \sigma\alpha^2 & \alpha \\
        \epsilon & \alpha & 1
    \end{pmatrix} e_{11}, \\
    d_i &= c_{t_i}.
\end{align}
Although we can provide an arbitrary degree sequence, for simplicity, we assume the nodes belonging to the same block $k$ have equal degree $c_k$. As in the canonical SBM, the model parameters must satisfy  constraint (\ref{eq:constraint_between_alp_eps}).

\subsection{Derivation of the spectrum and the detectability limit of the microcanonical SBM}
\label{sec:appendix_derivation_micro}
Here, we conduct an analysis analogous to Appendix~\ref{sec:appendix_b} for the microcanonical SBM. As a result of the present analysis, we obtain the same average largest eigenvalues as those of the canonical case in (\ref{eq:can_larget_eigenvalue_det}) and (\ref{eq:can_largest_eigenvalue_und}). However, a different technique is required to impose the microcanonical constraints. The calculations in this appendix are extensions of those in Refs.~\cite{kawamoto2015limitations,kawamoto2015detectability}. We start with the $n$th moment of the partition function (\ref{eq:can_nth_moment_Z})
\begin{align}
    [Z^n(M,\beta)]_M &=\int \left(\prod_{a=1}^nd\boldsymbol{x}_a\delta(\boldsymbol{x}_a^\top\boldsymbol{x}_a-N)\right)\nonumber\\
    &\quad\times\left[\exp{\left(\frac{\beta}{2}\sum_a\boldsymbol{x}_a^\top M\boldsymbol{x}_a\right)}\right]_M.
    \label{eq:micro_nth_moment_Z}
\end{align}
As defined in Appendix~\ref{sec:appendix_micro_model_definition}, we assume the three blocks model. Then, the exponential factor in (\ref{eq:micro_nth_moment_Z}) can be recast as
\begin{align}
    &\boldsymbol{x}_a^\top M\boldsymbol{x}_a \nonumber\\
    &=\sum_{ij\in V_1}u_{ij}x_{ia}x_{ja}+\sum_{ij\in V_2}y_{ij}x_{ia}x_{ja}+\sum_{ij\in V_3}u_{ij}x_{ia}x_{ja} \nonumber \\
    &\quad+2\sum_{i\in V_1}\sum_{j\in V_2}v_{ij}x_{ia}x_{ja}+2\sum_{i\in V_2}\sum_{j\in V_3}v_{ij}x_{ia}x_{ja} \nonumber\\
    &\quad+2\sum_{i\in V_1}\sum_{j\in V_3}w_{ij}x_{ia}x_{ja}-(\boldsymbol{\gamma}^\top\boldsymbol{x}_a)^2,
    \label{eq:expand_xmx}
\end{align}
where $u_{ij}, y_{ij}, v_{ij}$, and $ w_{ij}$ are the adjacency matrix elements. These parameters were introduced to distinguish blocks that obey different statistics.
Again, the summation $\sum_{i\in V_k}$ is taken over indices of the nodes that belong to block $k$. 

To calculate the ensemble average over the microcanonical SBM, we take the sum over all possible graph configurations as imposing the microcanonical constraints by delta functions.
Thus, the configuration average of the exponential factor in (\ref{eq:micro_nth_moment_Z}) is
\begin{widetext}
\begin{align}
    &\left[\exp{\left(\frac{\beta}{2}\sum_a\boldsymbol{x}_a^\top M\boldsymbol{x}_a\right)}\right]_M \nonumber \\
    &= \frac{1}{\mathcal{N}_G}\sum_{\{u_{ij}\},\{w_{ij}\},\{v_{ij}\},\{w_{ij}\}}
    \prod_{i\in V_1}\delta\left(\sum_{l\in V_1}u_{il}+\sum_{m\in V_2}v_{im}+\sum_{n\in V_3}w_{in}-c_1\right)
    \prod_{j\in V_2}\delta\left(\sum_{l\in V_1}u_{jl}+\sum_{m\in V_2}v_{jm}+\sum_{n\in V_3}w_{jn}-c_2\right)\nonumber \\
    &\times\prod_{k\in V_3}\delta\left(\sum_{l\in V_1}u_{kl}+\sum_{m\in V_2}v_{km}+\sum_{n\in V_3}w_{kn}-c_3\right)
    \delta\left(\sigma p_2\sum_{i\in V_1}\sum_{j\in V_2}v_{ij}-p_1\sum_{i,j\in V_2}y_{ij}\right)
    \delta\left(\sigma p_2\sum_{i\in V_2}\sum_{j\in V_3}v_{ij}-p_3\sum_{i,j\in V_2}y_{ij}\right)\nonumber \\
    &\times
    \delta\left(p_2\sum_{i,j\in V_1}u_{ij}-p_1\sum_{i\in V_1}\sum_{j\in V_2}v_{ij}\right)
    \delta\left(p_2\sum_{i,j\in V_3}u_{ij}-p_3\sum_{i\in V_2}\sum_{j\in V_3}v_{ij}\right) \nonumber \\
    &\times
    \delta\left(\epsilon\sum_{i,j\in V_1}u_{ij}-\sum_{i\in V_1}\sum_{j\in V_3}w_{ij}\right)
    \delta\left(\epsilon\sum_{i,j\in V_3}u_{ij}-\sum_{i\in V_1}\sum_{j\in V_3}w_{ij}\right)\exp{\left(\frac{\beta}{2}\sum_a\boldsymbol{x}_a^\top M\boldsymbol{x}_a\right)}.
    \label{eq:micro_exp_delta_form}
\end{align}
\end{widetext}
Here, $\mathcal{N}_G$ is the number of all realizable graphs that satisfy the constraints. The first three delta functions in (\ref{eq:micro_exp_delta_form}) represent Kronecker's deltas that impose the degree constraints, while the remaining ones represent Dirac's deltas that impose the constraints with respect to the number of edges between blocks, as specified by matrix $\boldsymbol{e}$.

We use the integral expression of the delta functions as follows.
\begin{align}
    \delta(x) &= \oint\frac{dz}{2\pi}z^{x-1},\label{eq:micro_delta_trans_z} \\
    \delta(x) &= \int_{-i\infty}^{i\infty}\frac{d\eta}{2\pi}e^{-\eta x}.
    \label{eq:micro_delta_trans_fourie}
\end{align}
Here, Eqs.~(\ref{eq:micro_delta_trans_z}) and (\ref{eq:micro_delta_trans_fourie}) correspond to the Kronecker's and Dirac's deltas.
Then, Eq.~(\ref{eq:micro_exp_delta_form}) can be recast as follows.
\begin{widetext}
\begin{align}
    &\frac{1}{\mathcal{N}_G}
    \oint\prod_{k=1,2,3}\prod_{i\in V_k}\frac{dz_i}{2\pi}z_i^{-(1+c_k)}\int\frac{d\zeta}{2\pi}\int\frac{d\xi}{2\pi}\int\frac{d\tau}{2\pi}\int\frac{d\kappa}{2\pi}\int\frac{d\eta}{2\pi}\int\frac{d\theta}{2\pi} e^{-\frac{\beta}{2}\sum_a(\boldsymbol{\gamma}^\top\boldsymbol{x}_a)^2}  \nonumber\\
    &\times
    \prod_{\substack{i<j\\i,j\in V_1}}\sum_{u_{ij}\in\{0,1\}}\left(z_iz_je^{\beta \sum_ax_{ia}x_{ja}-2\tau p_2-2\eta\epsilon}\right)^{u_{ij}}
    \prod_{\substack{i<j \\ i,j\in V_2}}\sum_{y_{ij}\in\{0,1\}}\left(z_iz_je^{\beta \sum_ax_{ia}x_{ja}+2\xi p_3+2\zeta p_1}\right)^{y_{ij}}
    \nonumber \\
    &\times
    \prod_{\substack{i<j \\ i,j\in V_3}}\sum_{u_{ij}\in\{0,1\}}\left(z_iz_je^{\beta \sum_ax_{ia}x_{ja}-2\kappa p_2-2\theta\epsilon}\right)^{u_{ij}}
    \prod_{i\in V_1}\prod_{j\in V_2}\sum_{v_{ij}\in\{0,1\}}\left(z_iz_je^{\beta \sum_ax_{ia}x_{ja}-\sigma\zeta p_2+\tau p_1}\right)^{v_{ij}}
    \nonumber \\
    &\times
    \prod_{i\in V_2}\prod_{j\in V_3}\sum_{v_{ij}\in\{0,1\}}\left(z_iz_je^{\beta \sum_ax_{ia}x_{ja}-\sigma\xi p_2+\kappa p_3}\right)^{v_{ij}}
    \prod_{i\in V_1}\prod_{j\in V_3}\sum_{w_{ij}\in\{0,1\}}\left(z_iz_je^{\beta \sum_ax_{ia}x_{ja}+\eta+\theta}\right)^{w_{ij}},\label{eq:after_insterting_xmx}
\end{align}
\end{widetext}
where parameters $\zeta, \xi, \tau, \kappa, \eta$, and $ \theta$ are the auxiliary variables provided by the integral representation of the delta function.
Because variables $u_{ij}, y_{ij}, v_{ij}$, and $ w_{ij}$ only take binary values, their summations in (\ref{eq:after_insterting_xmx}) can be calculated straightforwardly. For example,
\begin{align}
    &\prod_{\substack{i<j \\ i,j\in V_1}}\sum_{u_{ij}\in\{0,1\}}\left(z_iz_je^{\beta \sum_ax_{ia}x_{ja}}\right)^{u_{ij}} \nonumber\\
    &=\prod_{\substack{i<j\\i,j\in V_2}}(1+z_iz_je^{\beta \sum_ax_{ia}x_{ja}})\approx\prod_{\substack{i<j \\ i,j\in V_3}}\exp(z_iz_je^{\beta \sum_ax_{ia}x_{ja}}).
    \label{eq:micro_after_summation}
\end{align}
To derive the last equation in (\ref{eq:micro_after_summation}), we assume that $|z_i|$ and $|z_j|$ are sufficiently small.

Here, we introduce the order-parameter functions
\begin{equation}
    Q_k(\boldsymbol{\mu})=
    \frac{1}{p_kN}\sum_{i\in V_k}z_i\prod_{a=1}^n\delta(x_{ia}-\mu_a),
    \ \ \ (k=1,2,3)
    \label{eq:micro_orderparam_func}
\end{equation}
which is similar but not completely equivalent to (\ref{eq:app_order_parameter_function}).
Using the order-parameter functions (\ref{eq:micro_orderparam_func}), when $N\gg1$, Eq.~(\ref{eq:micro_after_summation}) can be approximated as
\begin{align}
    &\prod_{\substack{i<j\\i,j\in V_1}}\exp(z_iz_je^{\beta\sum_a x_{ia}x_
    {ja}}) \nonumber\\
    &\approx
    \exp\left(\frac{(p_1N)^2}{2}\int\prod_{a=1}^nd\mu_ad\nu_aQ_1(\boldsymbol{\mu})Q_1(\boldsymbol{\nu})e^{\beta\sum_a\mu_a\nu_a} \right),
    \label{eq:mincro_after_orderparam}
\end{align}
where we approximated that the contribution from the diagonal elements is negligible.
Using the similar calculations, (\ref{eq:micro_nth_moment_Z}) is now written as
\begin{equation}
    [Z^n(M,\beta)]_M=e^{N\mathcal{T}_n({Q})+N\mathcal{S}_n},
    \label{eq:zn_nt_ns}
\end{equation}
where 
\begin{align}
    &N\mathcal{T}_n({Q})\nonumber\\
    &=
    \frac{(p_1N)^2}{2}\int\prod_{a=1}^nd\mu_ad\nu_a{Q}_1(\boldsymbol{\mu}){Q}_1(\boldsymbol{\nu})e^{\beta\sum_a\mu_a\nu_a-2\tau p_2-2\eta\epsilon}
    \nonumber\\
    &+
    \frac{(p_2N)^2}{2}\int\prod_{a=1}^nd\mu_ad\nu_a{Q}_2(\boldsymbol{\mu}){Q}_2(\boldsymbol{\nu})e^{\beta\sum_a\mu_a\nu_a+2\xi p_3+2\zeta p_1}
    \nonumber\\
    &+
    \frac{(p_3N)^2}{2}\int\prod_{a=1}^nd\mu_ad\nu_a{Q}_3(\boldsymbol{\mu}){Q}_3(\boldsymbol{\nu})e^{\beta\sum_a\mu_a\nu_a-2\kappa p_2-2\theta\epsilon}
    \nonumber\\
    &+
    p_1p_2N^2\int\prod_{a=1}^nd\mu_ad\nu_a{Q}_1(\boldsymbol{\mu}){Q}_2(\boldsymbol{\nu})e^{\beta\sum_a\mu_a\nu_a-\sigma\zeta p_2+2\tau p_1}
    \nonumber\\
    &+
    p_2p_3N^2\int\prod_{a=1}^nd\mu_ad\nu_a{Q}_2(\boldsymbol{\mu}){Q}_3(\boldsymbol{\nu})e^{\beta\sum_a\mu_a\nu_a-\sigma\xi p_2+2\kappa p_3}
    \nonumber\\
    &+
    p_1p_3N^2\int\prod_{a=1}^nd\mu_ad\nu_a{Q}_1(\boldsymbol{\mu}){Q}_3(\boldsymbol{\nu})e^{\beta\sum_a\mu_a\nu_a+\eta+\theta}
    \label{eq:eNTn}
\end{align}
and
\begin{align}
    e^{N\mathcal{S}_n}&=\int \prod_{i=1}^N\prod_{a=1}^ndx_{ia}\prod_{a=1}^n\delta\left(\sum_{i=1}^Nx_{ia}^2-N\right) \nonumber\\
    &\quad\times\int\sqrt{N}\prod_{a=1}^nd\Omega_a\delta\left(\sqrt{N}\Omega_a-\sum_i\gamma_ix_{ia}\right)e^{-\frac{\beta}{2}\Omega_a^2}\nonumber \\
    &\quad\times\frac{1}{\mathcal{N}_G}
    \oint\prod_{k=1,2,3}\prod_{i\in V_k}\frac{dz_i}{2\pi}z_i^{-(1+c_k)} \nonumber\\
    &\quad\times\int\frac{d\zeta}{2\pi}\int\frac{d\xi}{2\pi}\int\frac{d\tau}{2\pi}\int\frac{d\kappa}{2\pi}\int\frac{d\eta}{2\pi}\int\frac{d\theta}{2\pi}.
    \label{eq:after_oderparam_omega}
\end{align}
Here, $\Omega_a$ is the order parameter defined in (\ref{eq:can_order_parameter_omega}).
As in the case of the canonical SBM in (\ref{eq:can_order_parameter_function_identity}),
for Eq.~(\ref{eq:after_oderparam_omega}), we insert the identity
\begin{align}
    1&=\prod_{k=1,2,3}p_kN\int\frac{D{Q}_k}{2\pi}
    \delta\left(\sum_{i\in V_k}z_i\prod_{a=1}^n\delta(x_{ia}-\mu_a)-p_kN{Q}_k(\boldsymbol{\mu})\right)\label{eq:micro_order_parameter_function_identity} \\
    &=
    \prod_{k=1,2,3}p_kN\int\frac{D{Q}_kD\hat{{Q}}_k}{2\pi} 
    \exp\left(\sum_{k=1,2,3}\int d\boldsymbol{\mu}\hat{{Q}}_k(\boldsymbol{\mu})\right. \nonumber\\
    &\qquad\times\left.\left(\sum_{i\in V_k}z_i\prod_{a=1}^n\delta(x_{ia}-\mu_a)-p_kN{Q}_k(\boldsymbol{\mu})\right)\right).
\end{align}
In (\ref{eq:micro_order_parameter_function_identity}), we perform the functional integration over the space of function ${Q}_k(\boldsymbol{\mu})$.
It is required to insert identity (\ref{eq:micro_order_parameter_function_identity}), because it indicates that we performed the replacement of a function in (\ref{eq:micro_orderparam_func}) by ${Q}_k(\boldsymbol{\mu})$. 
Furthermore,
using the integral representation of the delta functions (\ref{eq:app_can_delta_function_omega}) and (\ref{eq:app_can_delta_function_phi}), we obtain
\begin{align}
    &e^{N\mathcal{S}_n}\nonumber\\
    &=
    \int\prod_{k=1,2,3}p_kN\frac{D{Q}_kD{\hat{Q}}_k}{2\pi}\int\prod_a\frac{\beta d\phi_a}{4\pi}\int\prod_a\frac{\beta Nd\Omega_a\hat{\Omega}_a}{2\pi}\nonumber\\
    &\quad\times\int\frac{d\zeta}{2\pi}\int\frac{d\xi}{2\pi}\int\frac{d\tau}{2\pi}\int\frac{d\kappa}{2\pi}\int\frac{d\eta}{2\pi}\int\frac{d\theta}{2\pi} \nonumber\\
    &\quad\times \exp
    \left(
    -\log N_\mathcal{G}-N\sum_{k}K_k({Q}_k, {\hat{Q}}_k) \right.\nonumber\\
    &\qquad\left.+\frac{\beta N}{2}\sum_a(2\Omega_a\hat{\Omega}_a-\Omega_a^2+\phi_a)-\sum_k\log c_k! \right.\nonumber\\
    &\qquad\left.+\sum_{k=1,2,3}\log L_{k}\left(\hat{{Q}}_k, \{\hat{\Omega_a}\}, \{\phi_a\}\right)
    \right),
    \label{eq:e_nsn}
\end{align}
where
\begin{equation}
    K_k({Q}_k,\hat{{Q}}_k)=p_k\int d\boldsymbol{\mu}{Q}_k(\boldsymbol{\mu})\hat{{Q}}_k(\boldsymbol{\mu}), \label{eq:Kk}
\end{equation}
\begin{align}
    &L_{k}\left(\hat{{Q}}_k, \{\hat{\Omega_a}\}, \{\phi_a\}\right)=
    \int \prod_{i\in V_k}\prod_adx_{ia}\prod_{i\in V_k}
    \left(
    \hat{{Q}}_k^{c_k}(\boldsymbol{x}_i) \right.\nonumber\\
    &\quad\times\left.\exp\left(-\beta\sum_a\left(\sqrt{N}\hat{\Omega}_a\gamma_ix_{ia}+\frac{1}{2}\phi_ax_{ia}^2\right)\right)
    \right). \label{eq:Lk}
\end{align}
Here, we used the relation
\begin{align}
    &\oint\frac{dz_i}{2\pi}z_i^{-(1+c_k)}e^{z_i\hat{{Q}}_k(\boldsymbol{x}_i)} \nonumber\\
    &\quad=\oint\frac{dz_i}{2\pi}z_i^{-(1+c_k)}\sum_{m=0}^{\infty}\frac{1}{m!}\left(z_i\hat{{Q}}_k(\boldsymbol{x}_i)\right)^m \\
    &\quad=\sum_{m=0}^\infty\frac{1}{m!}\hat{{Q}}_k(\boldsymbol{x}_i)\oint\frac{dz_i}{2\pi}z_i^{m-(1+c_k)} \label{eq:micro_oint_zi_eziq_ud}\\
    &\quad=\frac{1}{c_k!}\hat{{Q}}_k^{c_k}(\boldsymbol{x}_i).
    \label{eq:micro_oint_zi_eziq}
\end{align}
Now, the variable depending on the node index $i$ only appears as $\boldsymbol{x}_i$. Hence, after the integral with respect to $\boldsymbol{x}_i$ is carried out in $L_{k}\left(\hat{{Q}}_k, \{\hat{\Omega_a}\}, \{\phi_a\}\right)$, Eq.~(\ref{eq:e_nsn}) can be expressed only with integrals over the auxiliary variables $\phi_a$, $\Omega_a$, $\hat{\Omega}_a$, $\zeta$, $\xi$, $\tau$, $\kappa$, $\eta$, $\theta$ and functional integrals over ${Q}_k(\boldsymbol{\mu})$ and $\hat{{Q}}_k(\boldsymbol{\mu})$.

For further calculations, as in the case of the canonical SBM (Eqs.~(\ref{eq:can_gaussian_mix}) and (\ref{eq:can_gaussian_mix_hat})), we assume the functional form of ${Q}_k$ and $\hat{{Q}}_k$ are restricted to the Gaussian mixtures as follows.
\begin{align}
    {Q}_k(\boldsymbol{\mu})&=T_k\int dAdHq_k(A,H) \nonumber\\
    &\quad\times\left(\frac{\beta A}{2\pi}\right)^{\frac{n}{2}}\exp\left(-\frac{\beta A}{2}\sum_a\left(\mu_a-\frac{H}{A}\right)^2\right), \label{eq:gaussian_mix} \\
    \hat{{Q}}_k(\boldsymbol{\mu})&=\hat{T}_k\int d\hat{A}d \hat{H}\hat{q}_k(\hat{A}, \hat{H})\exp \left(\frac{\beta}{2}\sum_a\left(\hat{A}\mu_a^2+2\hat{H}\mu_a\right)\right),
    \label{eq:gaussian_mix_hat}
\end{align}
where $T_k$ and $\hat{T}_k$ represent the normalization constants. With these functional forms, we can calculate the integrals over $\boldsymbol{\mu}$ in (\ref{eq:Kk}) and $\boldsymbol{x}$ in (\ref{eq:Lk}). Then, we obtain the following expressions.
\begin{align}
    &K_k(q_k,\hat{q}_k) \nonumber\\
    &= c_kp_k\int dAdH\int d\hat{A}d\hat{H}q_k(A,H)\hat{q}_k(\hat{A}, \hat{H})\nonumber\\
    &\quad\times\left(\frac{A}{A-\hat{A}}\right)^{\frac{n}{2}}\exp\left(\frac{n\beta}{2}\left(\frac{(H+\hat{H})^2}{A-\hat{A}}-\frac{H^2}{A}\right)\right)
    , \label{eq:Kk_gm}
\end{align}
\begin{align}
    &L_{k}\left(\hat{q}_k, \{\hat{\Omega_a}\}, \{\phi_a\}\right)=
    \hat{T}_k^{c_k}\left(\frac{2\pi}{\beta}\right)^{\frac{n}{2}} \nonumber\\
    &\quad\times\int \prod_{g=1}^{c_k}(d\hat{A}_gd\hat{H}_g\hat{q}_k(\hat{A}_g,\hat{H}_g))\prod_{a=1}^n \left(\phi_a-\sum_{g=1}^{c_k}\hat{A}_g\right)^{-\frac{1}{2}}\nonumber\\
    &\quad\times
    \exp \left(\frac{\beta}{2}\sum_{i\in V_k}\frac{(\sqrt{N}\hat{\Omega}_a\gamma_i-\sum_{g=1}^{c_k}\hat{H}_g)^2}{\phi_a-\sum_{g=1}^{c_k}\hat{A}_g}\right).
    \label{eq:Lk_gm}
\end{align}
In Appendix~\ref{sec:appendix_A}, we solve for normalization constants $T_k$ and $\hat{T}_k$. By using (\ref{eq:appendA_saddle_conditions9}), we can replace $T_k\hat{T}_k$ with $c_k$. This is how we eliminated the normalization constants in Eq.~(\ref{eq:Kk_gm}).
By inserting (\ref{eq:gaussian_mix}) and (\ref{eq:gaussian_mix_hat}) in (\ref{eq:eNTn}), we can calculate the integrals over $\boldsymbol{\mu}$ and obtain
\begin{align}
    &\mathcal{T}_n\nonumber\\
    &=N\int dAdH\int dA'dH'\left(\frac{AA'}{A-A'}\right)^\frac{n}{2}
    \exp\left(\frac{n\beta}{2}\Xi(A,A',H,H')\right) \nonumber \\
    & \quad\times\left( \frac{c_1}{2}\frac{p_1^2}{p_1+p_2+\epsilon p_1}q_1(A,H)q_1(A',H') \right.\nonumber\\
    &\qquad\left.+\frac{c_2}{2}\frac{\sigma p_2^2}{p_1+\sigma p_2+p_3} q_2(A,H)q_2(A',H') \right. \nonumber \\
    &\qquad + \frac{c_3}{2}\frac{p_3^2}{p_3+p_2+\epsilon p_3}q_3(A,H)q_3(A',H') \nonumber\\
    & \qquad+c_2\frac{p_1p_2}{p_1+\sigma p_2+p_3}q_1(A,H)q_2(A',H') \nonumber \\
    & \qquad\left.+ c_2\frac{p_2p_3}{p_1+\sigma p_2+\sigma p_3}q_2(A,H)q_3(A',H') \right.\nonumber\\
    &\qquad\left.+ c_1\frac{\epsilon p_1^2}{p_1+p_2+\epsilon p_1}q_1(A,H)q_3(A',H') \right),
    \label{eq:micro_after_appendixA}
\end{align}
where 
\begin{equation}
    \Xi(A,A',H,H')=\frac{A'H^2+AH'^2+2HH'}{AA'-1}-\frac{H^2}{A}-\frac{H'^2}{A'}.
\end{equation}
Here, we used the relations between $T_1$, $T_2$, and $T_3$ (\ref{eq:appendA_sad1_from_A_1})--(\ref{eq:appendA_sad1_from_A_7}).
From the calculations so far, we have performed all the integrals over $\boldsymbol{z}$, $\boldsymbol{x}$, and $\boldsymbol{\mu}$. The functional integrals over ${Q}_k(\boldsymbol{\mu})$ and $\hat{{Q}}_k(\boldsymbol{\mu})$ in (\ref{eq:e_nsn}) have been replaced by the integral over the functions $q_k(A,H)$ and $\hat{q}_k(\hat{A}, \hat{H})$.
In summary, the $n$th moment of the partition function (\ref{eq:zn_nt_ns}) is now represented by the integrals with respect to auxiliary variables $\phi_a$, $\Omega_a$, and $\hat{\Omega}_a$ and the functional integrals over $q_k(A,H)$ and $\hat{q}_k(\hat{A},\hat{H})$. Note that the other variables $\zeta$, $\xi$, $\tau$, $\kappa$, $\eta$, and $\theta$ can be erased when inserting the relations between the normalization constants (\ref{eq:appendA_sad1_from_A_1})--(\ref{eq:appendA_sad1_from_A_7}).

Again, as we assumed in the canonical SBM, we impose the replica symmetric assumptions for the parameters $\phi_a, \Omega_a$, and $\hat{\Omega}_a$, i.e., $\phi_a=\phi$, $\Omega_a=\Omega$, and $\hat{\Omega}_a=\hat{\Omega}$ in Eq.~(\ref{eq:Kk_gm})--(\ref{eq:micro_after_appendixA}). Inserting Eq.~(\ref{eq:Kk_gm})--(\ref{eq:micro_after_appendixA}) under the assumptions into (\ref{eq:zn_nt_ns}) and taking the limit $N\to\infty$, the average largest eigenvalue can be expressed as follows.
\begin{widetext}
\begin{align}
    &[\lambda(M)]_M \nonumber \\
    &\quad= 2\lim_{\beta\to\infty}\frac{1}{\beta N}\lim_{n\to0}\frac{\partial}{\partial n}\log [Z^n]_M \\
    &\quad=\underset{q_k,\hat{q}_k, \phi, \Omega,\hat{\Omega}}{\operatorname{extr}}
    \left\{
    \int dAdH\int dA'dH'\Xi(A,A',H,H')
    \right. \nonumber\\
    &\qquad\times\left( \frac{c_1}{2}\frac{p_1^2}{p_1+p_2+\epsilon p_1}q_1(A,H)q_1(A',H')+\frac{c_2}{2}\frac{\sigma p_2^2}{p_1+\sigma p_2+p_3} q_2(A,H)q_2(A',H') \right.+ \frac{c_3}{2}\frac{p_3^2}{p_3+p_2+\epsilon p_3}q_3(A,H)q_3(A',H') \nonumber\\
    &\qquad+c_2\frac{p_1p_2}{p_1+\sigma p_2+p_3}q_1(A,H)q_2(A',H') \left.+ c_2\frac{p_2p_3}{p_1+\sigma p_2+ p_3}q_2(A,H)q_3(A',H')
    + c_1\frac{\epsilon p_1^2}{p_1+p_2+\epsilon p_1}q_1(A,H)q_3(A',H') \right) \nonumber \\
    &\qquad -\sum_{k=1,2,3}c_kp_k\int dAdH\int d\hat{A}d\hat{H}q_k(A,H)\hat{q}_k(\hat{A},\hat{H})\left(\frac{(H+\hat{H})^2}{A-\hat{A}}-\frac{H^2}{A}\right)\nonumber \\
    &\qquad+ 2\Omega\hat{\Omega}-\Omega^2+\phi\nonumber \\
    &\qquad+\left. \frac{1}{N}\sum_{k=1,2,3}\int\prod_{g=1}^{c_k}
    \left(d\hat{A}_gd\hat{H}_g\hat{q}_k(\hat{A}_g,\hat{H}_g)\right)
    \sum_{i\in V_k}\frac{\left(\sqrt{N}\hat{\Omega}\gamma_i-\sum_{g=1}^{c_k}\hat{H}_g\right)^2}{\phi-\sum_{g=1}^{c_k}\hat{A}_g} \right\}.
    \label{eq:micro_functional_lam}
\end{align}
From Eq.~(\ref{eq:micro_functional_lam}), we obtain the saddle-point conditions as
\begin{align}
    \hat{q}_1(\hat{A},\hat{H})&=
    \int dA'dH'\frac{p_1q_{1}(A',H')+p_2q_{2}(A',H')+\epsilon p_1q_{3}(A',H')}{p_1+p_2+\epsilon p_1}
    \delta\left(\hat{A}-\frac{1}{A'}\right)
    \delta\left(\hat{H}-\frac{H'}{A'}\right),\label{eq:micro_func_saddle_equation1}\\
    \hat{q}_2(\hat{A},\hat{H})&=
    \int dA'dH'\frac{p_1q_{1}(A',H')+\sigma p_2q_{2}(A',H')+ p_3q_{3}(A',H')}{p_1+\sigma p_2+p_3}
    \delta\left(\hat{A}-\frac{1}{A'}\right)
    \delta\left(\hat{H}-\frac{H'}{A'}\right),\label{eq:micro_func_saddle_equation2}\\
    \hat{q}_3(\hat{A},\hat{H})&=
    \int dA'dH'\frac{p_3q_{1}(A',H')+p_2q_{2}(A',H')+ \epsilon p_3q_{3}(A',H')}{p_3+ p_2+\epsilon p_3}
    \delta\left(\hat{A}-\frac{1}{A'}\right)
    \delta\left(\hat{H}-\frac{H'}{A'}\right),
    \label{eq:micro_func_saddle_equation3}
\end{align}
\end{widetext}
and
\begin{align}
    &q_k(A,H)=\frac{1}{p_kN}\int\prod_{g=1}^{c_k-1}
    \left(d\hat{A}_gd\hat{H}_g\hat{q}_k(\hat{A}_g,\hat{H}_g)\right) \nonumber\\
    &\times\delta\left(H-\sum_{g=1}^{c_k-1}\hat{H}_g+\sqrt{N}\hat{\Omega}\gamma_i\right)
    \delta\left(A-\phi+\sum_{g=1}^{c_k-1}\hat{A}_g\right).
    \label{eq:micro_func_saddle_equation_hat}
\end{align}
Moreover, the saddle-point conditions with respect to $\phi$ yield
\begin{equation}
    \sum_kp_k\int dAdH\mathcal{Q}_k(A,H)\left(\frac{H}{A}\right)^2=1,
    \label{eq:micro_func_constraint}
\end{equation}
where 
\begin{align}
    &\mathcal{Q}_k(A,H)=\frac{1}{p_kN}\sum_{i\in V_k}\int\prod_{g=1}^{c_k}
    \left(d\hat{A}_gd\hat{H}_{g}\hat{q}_k(\hat{A}_g,\hat{H}_g)\right) \nonumber\\
    &\times\delta\left(H-\sum_{g=1}^{c_k}\hat{H}_g+\sqrt{N}\hat{\Omega}\gamma_i\right)
    \delta\left(A-\phi+\sum_{g=1}^{c_k}\hat{A}_g\right).
\end{align}
Equations~(\ref{eq:micro_func_saddle_equation1})--(\ref{eq:micro_func_saddle_equation_hat}) constitute functional equations under the constraint (\ref{eq:micro_func_constraint}). This constraint corresponds to the normalization constraints in (\ref{eq:modularity_maximization_problem2}).
By solving these equations, we obtain the distribution of the largest eigenvector elements.

As in the canonical case, solving the functional form of equations is still not analytically tractable. Thus, we again introduce the EMA, i.e., the precision parameters of the Gaussian mixtures $A$ and $\hat{A}$ are fixed as constants, i.e., $q_k(A,H)=q(H)\delta(A-a_k)$ and $\hat{q}_k(\hat{A},\hat{H})=\hat{q}_k(\hat{H})\delta(\hat{A}-\hat{a}_k)$. Performing the EMA for (\ref{eq:micro_functional_lam}), we have
\begin{widetext}
\begin{align}
    [\lambda(M)]_M=&\ \underset{\phi, \Omega,\hat{\Omega},m_{1k}, m_{2k},\hat{m}_{1k}, \hat{m}_{2k}, a_k,\hat{a}_k}{\operatorname{extr}}\left[
    \frac{c_1p_1^2}{p_1+p_2+\epsilon p_1}\frac{a_1m_{21}+m_{11}^2}{a_1^2-1}+
    \frac{c_2\sigma p_2^2}{p_1+\sigma p_2+p_3}\frac{a_2m_{22}+m_{12}^2}{a_2^2-1}\right. \nonumber \\
    &+ \frac{c_3p_3^2}{p_3+p_2+\epsilon p_3}\frac{a_3m_{23}+m_{13}^2}{a_3^2-1} 
    + \frac{c_2p_1p_2}{p_1+\sigma p_2+p_3}\frac{a_2m_{21}+a_1m_{22}+2m_{11}\hat{m}_{11}}{a_1a_2-1} \nonumber \\
    &+ \frac{c_2p_2p_3}{p_1+\sigma p_2+p_3}\frac{a_3m_{22}+a_2m_{23}+2m_{12}m_{13}}{a_2a_3-1}+\frac{c_1\epsilon p_1^2}{p_1+p_2+\epsilon p_1}\frac{a_3m_{21}+a_1m_{23}+2m_{11}m_{13}}{a_1a_3-1} \nonumber \\
    &-\sum_kc_kp_k\frac{m_{2k}+2m_{1k}\hat{m}_{1k}+\hat{m}_{2k}}{a_k-\hat{a}_k} + 2\Omega\hat{\Omega}-\Omega^2+\phi \nonumber \\
    &+ \left.\frac{1}{N}\sum_k\sum_{i\in V_k}\frac{1}{\phi-c_k\hat{a}_k}\left((\sqrt{N}\hat{\Omega}\gamma_i)^2-2\sqrt{N}\hat{\Omega}\gamma_ic_k\hat{m}_{1k}+c_k\hat{m}_{2k}+c_k(c_k-1)\hat{m}_{1k}^2\right)
    \right],
    \label{eq:micro_lambda_after_ema}
\end{align}
\end{widetext}
where $m_{\ell k}$ and $\hat{m}_{\ell k}$ represent the $\ell$th moments of $H$ and $\hat{H}$, respectively, i.e., $m_{\ell k}=\int dHH^\ell q_k(H)$ and $\hat{m}_{\ell k}=\int d\hat{H}\hat{H}^\ell \hat{q}_k(\hat{H})$.

As in the canonical case, we introduce further assumptions. First, we assume the symmetry between the community blocks, namely $p_1=p_3$ and $c_1=c_3$. Hence, $a_1=a_3$, $\hat{a}_1=\hat{a}_3$, $m_{21}=m_{23}$, and $\hat{m}_{21}=\hat{m}_{23}$.  Second, we think of two types of solutions: $m_{11}=-m_{13},\ m_{12}=0$ and $m_{11}=m_{12}=m_{13}=0$.
Under these assumptions, we obtain the same solutions as those of the canonical SBM with the regular approximation.
When $m_{11}=-m_{13}$ and $m_{12}=0$, the average largest eigenvalue is obtained as in Eq.~(\ref{eq:can_larget_eigenvalue_det}). 
When $m_{11}=m_{12}=m_{13}=0$, the average largest eigenvalue is obtained as in Eq.~(\ref{eq:can_largest_eigenvalue_und}). 
The detectability limit is given by Eq.~(\ref{eq:can_detectability_threshold}).

\subsection{Saddle-point conditions for $\mathcal{N}_G$}
\label{sec:appendix_A}
The goal of this subsection is to derive the relations of the normalization constants of the Gaussian mixtures $T_k$ and $\hat{T}_k$ in (\ref{eq:gaussian_mix}) and (\ref{eq:gaussian_mix_hat}). They can be derived using saddle-point conditions for the number of all realizable graphs $\mathcal{N}_G$. This can be calculated by taking the sum over all possible graph configurations as imposing the microcanonical constraints by delta functions. Thus, we have
\begin{widetext}
\begin{align}
    \mathcal{N}_G
    &=\sum_{\{u_{ij}\},\{w_{ij}\},\{v_{ij}\},\{w_{ij}\}}
    \prod_{i\in V_1}\delta\left(\sum_{l\in V_1}u_{il}+\sum_{m\in V_2}v_{im}+\sum_{n\in V_3}w_{in}-c_1\right) \nonumber\\
    &\quad\times\prod_{j\in V_2}\delta\left(\sum_{l\in V_1}u_{jl}+\sum_{m\in V_2}v_{jm}+\sum_{n\in V_3}w_{jn}-c_2\right)
    \prod_{k\in V_3}\delta\left(\sum_{l\in V_1}u_{kl}+\sum_{m\in V_2}v_{km}+\sum_{n\in V_3}w_{kn}-c_3\right) \nonumber\\
    &\quad\times\delta\left(\sigma p_2\sum_{i\in V_1}\sum_{j\in V_2}v_{ij}-p_1\sum_{i,j\in V_2}y_{ij}\right)
    \delta\left(\sigma p_2\sum_{i\in V_2}\sum_{j\in V_3}v_{ij}-p_3\sum_{i,j\in V_2}y_{ij}\right)\nonumber \\
    &\quad\times 
    \delta\left(p_2\sum_{i,j\in V_1}u_{ij}-p_1\sum_{i\in V_1}\sum_{j\in V_2}v_{ij}\right)
    \delta\left(p_2\sum_{i,j\in V_3}u_{ij}-p_3\sum_{i\in V_2}\sum_{j\in V_3}v_{ij}\right) \nonumber \\
    &\quad\times 
    \delta\left(\epsilon\sum_{i,j\in V_1}u_{ij}-\sum_{i\in V_1}\sum_{j\in V_3}w_{ij}\right)
    \delta\left(\epsilon\sum_{i,j\in V_3}u_{ij}-\sum_{i\in V_2}\sum_{j\in V_3}w_{ij}\right).
\end{align}
Using the integral representation of the delta function (\ref{eq:micro_delta_trans_z}) and (\ref{eq:micro_delta_trans_fourie}),
we have
\begin{align}
     \mathcal{N}_G&=\sum_{\{u_{ij}\},\{w_{ij}\},\{v_{ij}\},\{w_{ij}\}}
    \oint\prod_{i\in V_1}\frac{dz_i}{2\pi}z_i^{\sum_{l\in V_1}u_{il}+\sum_{m\in V_2}v_{im}+\sum_{n\in V_3}w_{in}-c_1-1} \nonumber \\
    &\quad\times\oint\prod_{i\in V_2}\frac{dz_i}{2\pi}z_i^{\sum_{l\in V_1}v_{il}+\sum_{m\in V_2}y_{im}+\sum_{n\in V_3}v_{in}-c_2-1}
    \oint\prod_{i\in V_3}\frac{dz_i}{2\pi}z_i^{\sum_{l\in V_1}u_{il}+\sum_{m\in V_2}v_{im}+\sum_{n\in V_3}w_{in}-c_3-1} \nonumber \\
    &\quad\times\int \frac{d\zeta}{2\pi}e^{-\zeta\left(\sigma p_2\sum_{i\in V_1}\sum_{j\in V_2}v_{ij}-p_1\sum_{i,j\in V_2}y_{ij}\right)}
    \int \frac{d\xi}{2\pi}e^{-\xi\left(\sigma p_2\sum_{i\in V_2}\sum_{j\in V_3}v_{ij}-p_3\sum_{i,j\in V_2}y_{ij}\right)} \nonumber\\
    &\quad\times \int \frac{d\tau}{2\pi}e^{-\tau\left( p_2\sum_{i,j\in V_1}u_{ij}-p_1\sum_{i\in V_1}\sum_{j\in V_2}v_{ij}\right)}
    \int \frac{d\kappa}{2\pi}e^{-\kappa\left( p_2\sum_{i,j\in V_1}u_{ij}-p_3\sum_{i\in V_2}\sum_{j\in V_3}v_{ij}\right)} \nonumber\\
    &\quad\times\int \frac{d\eta}{2\pi}e^{-\eta\left( \epsilon\sum_{i,j\in V_1}u_{ij}-\sum_{i\in V_1}\sum_{j\in V_3}w_{ij}\right)}
    \int \frac{d\theta}{2\pi}e^{-\theta\left( \epsilon\sum_{i,j\in V_3}u_{ij}-\sum_{i\in V_1}\sum_{j\in V_3}w_{ij}\right)}
    \label{eq:appendA_after_x_trans}
\end{align}

\begin{align}
    &=\oint\prod_{k=1,2,3}\prod_{i\in V_k}\frac{dz_i}{2\pi}z_i^{-(1+c_k)}\int\frac{d\zeta}{2\pi}\int\frac{d\xi}{2\pi}\int\frac{d\tau}{2\pi}\int\frac{d\kappa}{2\pi}\int\frac{d\eta}{2\pi}\int\frac{d\theta}{2\pi} \nonumber \\
    &\quad\times
    \prod_{\substack{i<j\\i,j\in V_1}}\sum_{u_{ij}}\left(z_iz_je^{-2\tau p_2-2\eta\epsilon}\right)^{u_{ij}}
    \prod_{\substack{i<j\\i,j\in V_2}}\sum_{y_{ij}}\left(z_iz_je^{ 2\xi p_3+2\zeta p_1}\right)^{y_{ij}}
    \prod_{\substack{i<j\\i,j\in V_3}}\sum_{u_{ij}}\left(z_iz_je^{ -2\kappa p_2-2\theta\epsilon}\right)^{u_{ij}} \nonumber\\
    &\quad\times
    \prod_{i\in V_1}\prod_{j\in V_2}\sum_{v_{ij}}\left(z_iz_je^{-\sigma\zeta p_2+\tau p_1}\right)^{v_{ij}}
    \prod_{i\in V_2}\prod_{j\in V_3}\sum_{v_{ij}}\left(z_iz_je^{-\sigma\xi p_2+\kappa p_3}\right)^{v_{ij}}
    \prod_{i\in V_1}\prod_{j\in V_3}\sum_{w_{ij}}\left(z_iz_je^{\eta+\theta}\right)^{w_{ij}}. \label{eq:appendA_after_insterting_xmx}
\end{align}
\end{widetext}

Here, we introduce the order parameters
\begin{equation}
    q_k=\frac{1}{p_kN}\sum_{i\in V_k}z_i.\ \ \ (k=1,2,3)
    \label{eq:appc_order_parameter_qk}
\end{equation}
Equation (\ref{eq:appendA_after_insterting_xmx}) is now written as
\begin{align}
    \mathcal{N}_G&=\prod_{k=1,2,3}\left(p_kN\int dq_k\prod_{i\in V_k}\oint\frac{dz_i}{2\pi}z_i^{-(1+c_k)}\right) \nonumber\\
    &\quad\times\int\frac{d\zeta}{2\pi}\int\frac{d\xi}{2\pi}\int\frac{d\tau}{2\pi}\int\frac{d\kappa}{2\pi}\int\frac{d\eta}{2\pi}\int\frac{d\theta}{2\pi}\nonumber \\
    &\quad\times\prod_{k=1,2,3}\delta\left(p_kNq_k-\sum_{i\in V_k}z_i\right)\nonumber \\
    &\quad\times\exp\left(\frac{1}{2}e^{-2\tau p_2-2\epsilon\eta}(p_1Nq_1)^2
    +\frac{1}{2}e^{2\zeta p_1+2\xi p_3}(p_2Nq_2)^2\right. \nonumber\\
    &\qquad \left.+\frac{1}{2}e^{-2\kappa p_2-2\xi\theta}(p_3Nq_3)^2 + e^{-\sigma\zeta p_2+\tau p_1}p_1p_2N^2q_1q_2 \right.\nonumber\\
    &\qquad\left.+e^{-\sigma\xi p_2+\kappa p_3}p_2p_3N^2q_2q_3+e^{\eta+\theta}p_1p_3N^2q_1q_3)\right).
    \label{eq:appendA_Ng_after_order_param}
\end{align}
Here, we used the same approximation as in (\ref{eq:micro_after_summation}).
Using relations (\ref{eq:micro_delta_trans_fourie}) and (\ref{eq:micro_oint_zi_eziq}), Eq.~(\ref{eq:appendA_Ng_after_order_param}) becomes
\begin{align}
    \mathcal{N}_G&=\prod_{k=1,2,3}\left(p_kN\int \frac{dq_kd\hat{q}_k}{2\pi}\right)\nonumber\\
    &\quad\times\int\frac{d\zeta}{2\pi}\int\frac{d\xi}{2\pi}\int\frac{d\tau}{2\pi}\int\frac{d\kappa}{2\pi}\int\frac{d\eta}{2\pi}\int\frac{d\theta}{2\pi}\nonumber \\
    &\quad\times\exp\left(\frac{1}{2}e^{-2\tau p_2-2\epsilon\eta}(p_1Nq_1)^2
    +\frac{1}{2}e^{2\zeta p_1+2\xi p_3}(p_2Nq_2)^2 \right.\nonumber\\
    &\qquad\left.+\frac{1}{2}e^{-2\kappa p_2-2\xi\theta}(p_3Nq_3)^2+e^{-\sigma\zeta p_2+\tau p_1}p_1p_2N^2q_1q_2 \right.\nonumber\\
    &\qquad\left.+e^{-\sigma\xi p_2+\kappa p_3}p_2p_3N^2q_2q_3+e^{\eta+\theta}p_1p_3N^2q_1q_3) \right.\nonumber\\
    & \qquad+\left. N\sum_{k=1,2,3}(-\hat{q}_kp_kq_k+p_kc_k\log\hat{q}_k-p_k\log c_k!)\right).
    \label{eq:appa_ng_before_limit}
\end{align}
In the limit $N\to\infty$, we have the following saddle-point conditions.
\begin{align}
    \epsilon p_1q_1e^{-2\tau p_2-2\epsilon\eta} &= p_3q_3e^{\eta+\theta}\label{eq:appendA_saddle_conditions1} \\
    \epsilon p_3q_3e^{-2\kappa p_2-2\epsilon\theta} &= p_1q_1e^{\eta+\theta} \\
    q_2e^{2\zeta p_1+2\xi p_3}&= \sigma q_1e^{-\sigma\zeta p_2+\tau p_1} \\
    q_1e^{-2\tau p_2-2\epsilon\eta}&=q_2e^{-\sigma\zeta p_2+\tau p_1} \\
    q_3e^{-2\kappa p_2-2\epsilon\theta}&=q_2e^{-\sigma\xi p_2+\kappa p_3} 
\end{align}
\begin{align}
    \frac{\hat{q}_1}{N}&=p_1q_1e^{-2\tau p_2-2\epsilon\eta}+p_2q_2e^{-\sigma\zeta p_2+\tau p_1}+p_3q_3e^{\eta+\theta} \\
    \frac{\hat{q}_2}{N}&=p_2q_2e^{-2\zeta p_1+2\xi p_3}+p_1q_1e^{-\sigma\zeta p_2+\tau p_1}+p_3q_3e^{-\sigma\xi p_2+\kappa p_3} \\
    \frac{\hat{q}_3}{N}&=p_3q_3e^{-2\kappa p_2-2\epsilon\theta}+p_2q_2e^{-\sigma\xi p_2+\kappa p_3}+p_1q_1e^{\eta+\theta} \\
    q_k\hat{q}_k &= c_k. \ \ \ (k=1,2,3) \label{eq:appendA_saddle_conditions9}
\end{align}
From Eq.~(\ref{eq:appendA_saddle_conditions1})--(\ref{eq:appendA_saddle_conditions9}), we obtain
\begin{align}
    q_1^2 &=\frac{1}{N}e^{2\tau p_2+2\epsilon\eta}\frac{c_1}{p_1+p_2+\epsilon p_1}, \label{eq:appendA_sad1_from_A_1} \\
    q_2^2 &= \frac{1}{N}e^{-2\zeta p_1-2\xi p_3}\frac{c_2\sigma}{p_1+\sigma p_2+ p_3}, \\
    q_3^2 &= \frac{1}{N}e^{2\kappa p_2+2\epsilon\theta}\frac{c_3}{p_3+p_2+\epsilon p_3}, \\
    q_1q_2 &= \frac{1}{N}e^{\sigma\zeta p_2-\tau p_1}\frac{c_2}{\sigma p_2+p_1+ p_3}, \\
    q_2q_3 &= \frac{1}{N}e^{\sigma\xi p_2-\kappa p_3}\frac{c_2}{\sigma p_2+p_1+ p_3}, \\
    q_1q_3 &= \frac{1}{N}e^{-(\eta+\theta)}\frac{p_1}{p_3}\frac{c_1\epsilon}{ p_1+p_2+ \epsilon p_1}, \label{eq:appendA_sad1_from_A_6} \\
    c_1(\sigma p_2+p_1+p_3) &= c_2(p_1+p_2+\epsilon p_3).
    \label{eq:appendA_sad1_from_A_7}
\end{align}
By substituting (\ref{eq:appendA_sad1_from_A_1})--(\ref{eq:appendA_sad1_from_A_7}) into (\ref{eq:appa_ng_before_limit}), $\mathcal{N}_G$ is expressed in terms of the model parameters. The order parameters (\ref{eq:appc_order_parameter_qk}) correspond to the order-parameter functions (\ref{eq:app_order_parameter_function}) when $n=0$. This indicates that the normalization constants of the Gaussian mixtures $T_k$ and $\hat{T}_k$ in (\ref{eq:gaussian_mix}) and (\ref{eq:gaussian_mix_hat}) are identical to $q_k$ and $\hat{q}_k$, respectively. Accordingly, we obtain the relations between $T_k$ and $\hat{T}_k$ as Eqs.~(\ref{eq:appendA_saddle_conditions9})--(\ref{eq:appendA_sad1_from_A_6}). Besides, (\ref{eq:appendA_sad1_from_A_7}) is identical to the constraint between the model parameters (\ref{eq:constraint_between_alp_eps}), i.e., the same constraint is derived by both the model definition and the replica analysis.

\subsection{Comparison between the canonical and microcanonical SBMs}
\label{sec:appendix_comparison_two_models}
In the main text, we used the canonical SBM for deriving the detectability limit, whereas we used the microcanonical SBM for conducting the numerical experiments.
This is because the derivation under the canonical SBM is more straightforward and simpler, while the canonical SBM causes a problem when conducting the numerical experiments. 
The canonical SBM required the regular approximation as an additional approximation to calculate the average largest eigenvalue in the replica analysis. The approximation creates a large difference of the derived solutions from the original ones because of ignoring the fluctuation of the degree distribution. Thus, it becomes difficult to validate the results of the analytical calculation by comparing them to the results of the numerical experiments.

However, the microcanonical SBM does not require the regular approximation because it can be defined with an arbitrary degree sequence, and we can choose one that avoids the effects of the fluctuation.
Meanwhile, as mentioned in Sec.~\ref{sec:detectability_phase_diagram_and_the_lleading_eigenvalue}, the microcanonical SBM requires an additional constraint that $c_1$ and $c_2$ can take only natural numbers. This originates from the fact that it specifies a certain degree for each node as its model parameters. 
Note that the replica analysis with the microcanonical SBM (and canonical SBM) required another approximation, which is called EMA. However, the effect of this approximation can be neglected under the experimental condition in Sec.~\ref{sec:numerical_experiments}, as discussed in Appendix~\ref{sec:appendix_approximation_accuracy}.

In short, the canonical SBM is appropriate to explain the derivation of the detectability limit because of the simplicity. The microcanonical SBM is appropriate for conducting the numerical experiments because it does not require the regular approximation.

\section{Bimodal stochastic block model}
\label{sec:appendix_bimodal}
In this appendix, we explain the bimodal SBM in detail. 
This model is a variant of the SBM that has no overlapping structure. The bimodal SBM has a bimodal degree distribution:
each node randomly takes either degree $c_1$ or $c_2$. We denote the fraction of the nodes that have degree $c_1$ as $b_1$ and that of $c_2$ as $b_2$ ($b_1+b_2=1$). Note that, because the degree assignment is independent of the group assignment, one cannot infer the planted structure based on the degree sequence.

We define the two-blcok bimodal SBM in the microcanonical formulation. The model is parametrized as follows.
\begin{align}
    \boldsymbol{e} &=
    \begin{pmatrix}
    1 & \epsilon \\
    \epsilon & 1 \\
    \end{pmatrix}e_{11}, \\
    \boldsymbol{b}&=(b_1, b_2)=(2p_1, p_2).
\end{align}
Here, as defined in Sec.~\ref{sec:overlapping_sbm}, $e_{kl}$ is the number of edges between blocks $k$ and $l$, and $\epsilon$ is the parameter that controls the strength of community structure. Moreover, $p_1$ and $p_2(=\alpha p_1)$ are the sizes of the community and overlapping blocks of the overlapping SBM, respectively. As mentioned in the main text,
the purpose of introducing the bimodal SBM is to compare the overlapping SBM to the SBM with the non-overlapping structure and the same average degree. We can confirm that both models have the same average degree.

Subsequently, we show the average largest eigenvalue of the bimodal SBM under the detectable and undetectable conditions. As in the overlapping SBM, we can calculate it using the replica method. The detailed derivation can be found in Ref.~\cite{kawamoto2015limitations}.

First, under the detectable condition, we obtain the equation for $a$ as
\begin{equation}
    \overline{c_b}(c_2A-B)(c_1A-B)=(a^2-1)\left(\overline{c_b}A-B\right)B,
    \label{eq:bimodal_a_eq_det}
\end{equation}
where
\begin{align}
    A&=(\overline{c_b}-1)\Gamma -a, \\
    B&=\Gamma(\overline{c^2}-\overline{c_b})-a\overline{c_b}, \\
    \Gamma&=\frac{1-\epsilon}{1+\epsilon}.
\end{align}
Here, $a$ is the precision parameter of the Gaussian mixture, which corresponds to $a_1$ and $a_2$ in the case of the overlapping SBM. Besides, $\overline{c_b}\equiv b_1c_1+b_2c_2$ and $\overline{c^2_b}\equiv b_1c_1^2+b_2c_2^2$. 
Note that $a$ has no indices because of the symmetry between the two blocks. We let the solutions of Eq.~(\ref{eq:bimodal_a_eq_det}) be $a^{\operatorname{det}}$. Using this solution, we obtain the following expression of the average largest eigenvalue.
\begin{equation}
    [\lambda(M)]_M=\frac{c_1c_2}{(a^{\operatorname{det}})^3}\frac{A}{B}.
\end{equation}

Second, under the undetectable case, we obtain the equations for $a$ and $\phi$ as follows.
\begin{equation}
    \sum_{t=1,2}\frac{b_tc_t^2}{(\phi-c_t/a)^2}
    =\frac{(a^2+1)}{\overline{c_b}}
    \left(\frac{\overline{c_b}a}{a^2-1}\right)^2. \label{eq:bimodal_a_eq}
\end{equation}
When we let the solutions of these equations be $a^{\operatorname{und}}$ and $\phi^{\operatorname{und}}$, we obtain the average largest eigenvalue as $[\lambda(M)]_M=\phi^{\operatorname{und}}$.

\section{Accuracies of the EMA and the regular approximation}
\label{sec:appendix_approximation_accuracy}
For the replica analysis, we introduced two approximations: the regular approximation and EMA. Here, we investigate the dependencies of the average degree on the accuracy of each approximation. It is known that when the average degree is sufficiently large, the effect of these approximations can be asymptotically ignored. However, it is not trivial how the approximations affect the results for a graph with a low average degree.

\begin{figure*}[thb!]
\centering
\begin{tabular}{c}

\begin{minipage}{0.45\hsize}
\centering
\includegraphics[clip, width=8.4cm]{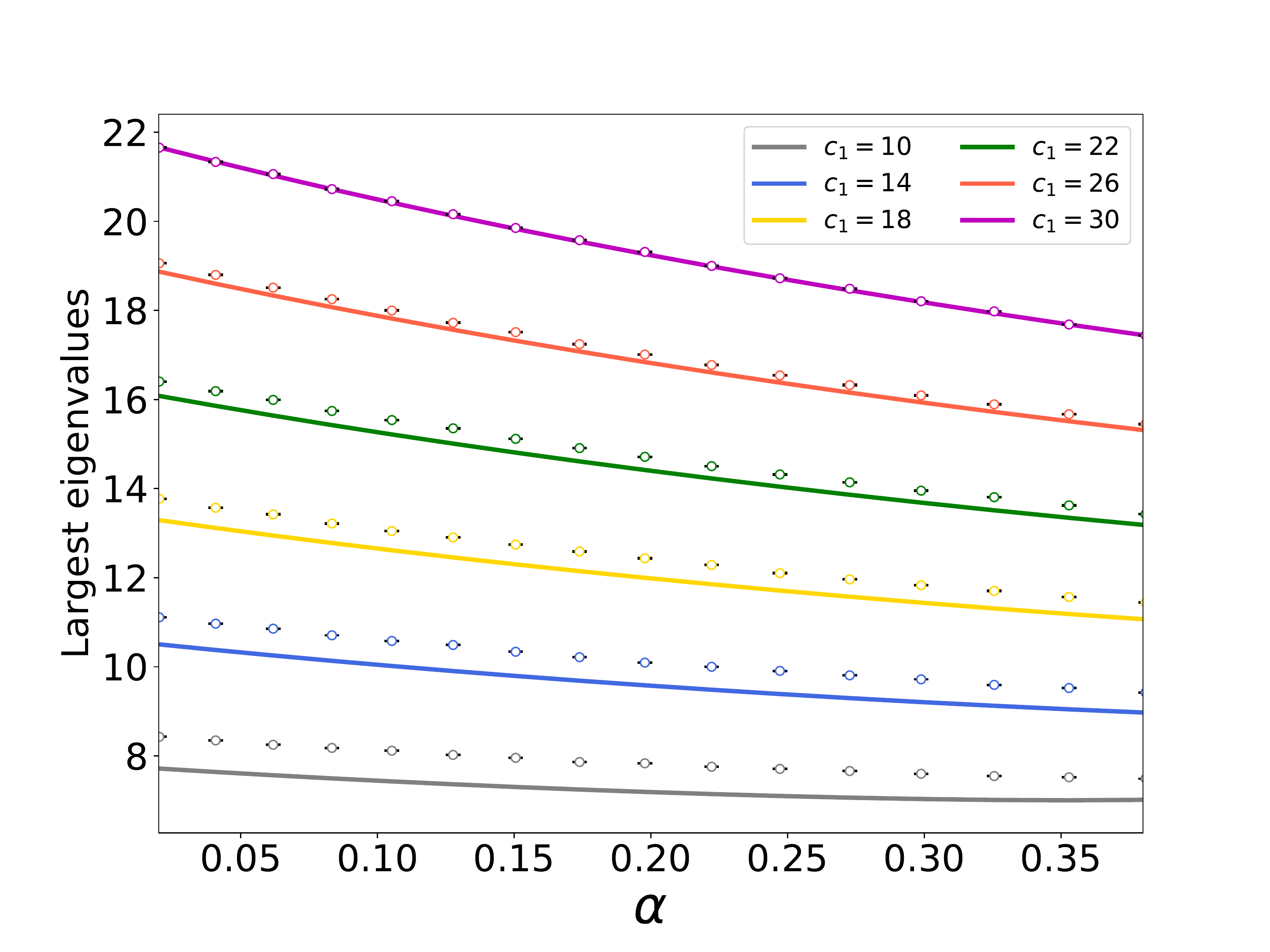}
\subcaption{}
\label{fig:regular_approx_performance}
\end{minipage}

\begin{minipage}{0.45\hsize}
\centering
\includegraphics[clip, width=8.4cm]{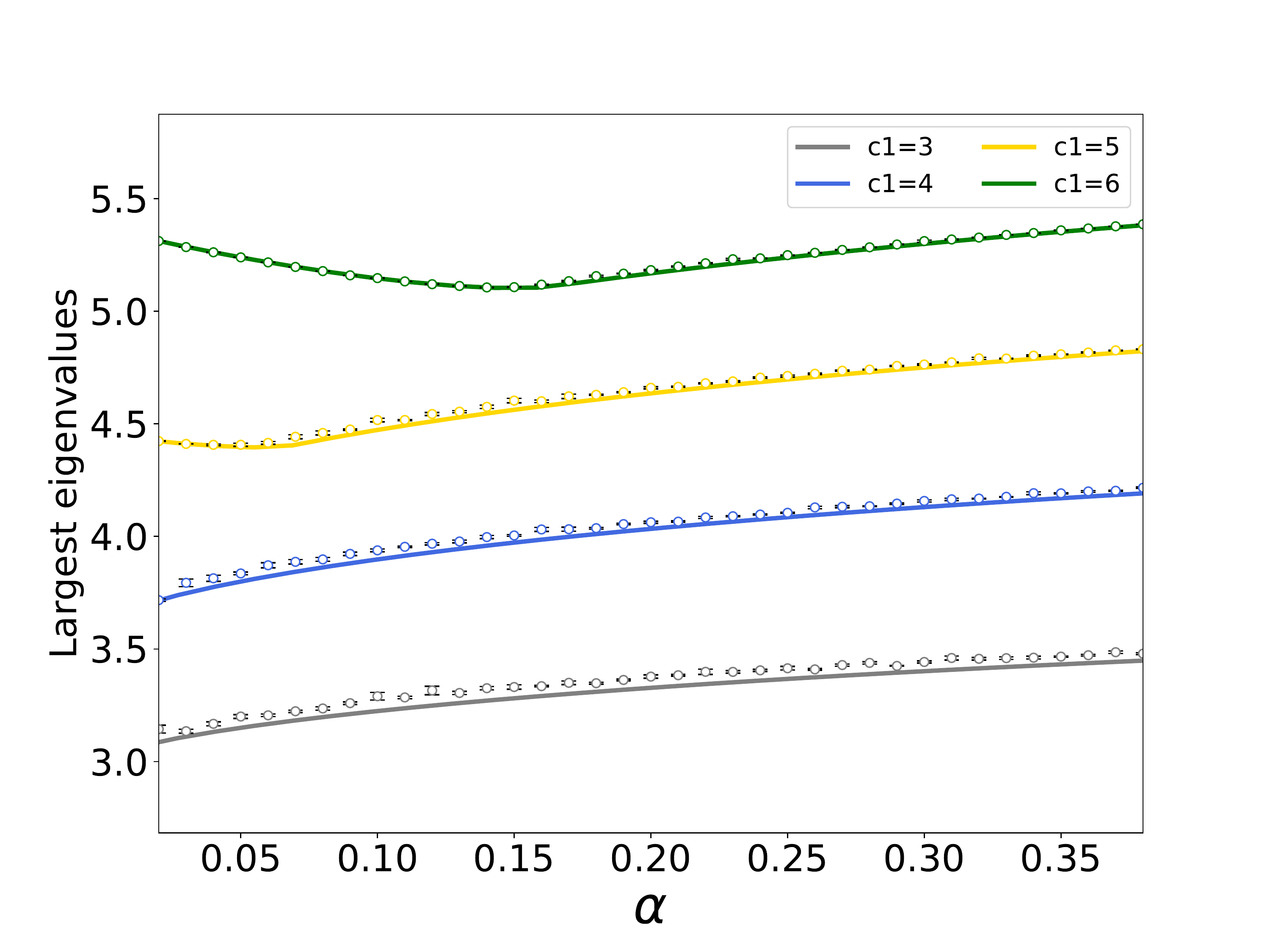}
\subcaption{}
\label{fig:ema_performance_c1_varies}
\end{minipage}

\end{tabular}
\caption{Largest eigenvalues as a function of $\alpha$. The lines represent the results of the replica analysis and the dots represent those of the numerical experiments. (a) The figure shows the results of the canonical SBM for $c_1=10,14,18,22,26,30$. (b) The figure shows the results of the microcanonical SBM for $c_1=3,4,5,6$.}
\label{fig:approximation_influence}
\end{figure*}

To derive the detectability limit of the canonical SBM, we used both the EMA and the regular approximation. To derive that of the microcanonical SBM, we used the EMA only. Thus, by comparing both results, we can measure how each approximation differs from the original result. Figs.~\ref{fig:regular_approx_performance} and \ref{fig:ema_performance_c1_varies} show the results of the canonical and microcanonical SBMs, respectively. 
We can see that the results of the replica analysis and the numerical experiments are in agreement for $c_1\geq 30$ in the canonical case. On the other hand, they are in agreement for $c_1\geq 6$ in the microcanonical case. Therefore, we can conclude that the effect of the EMA is smaller than that of the regular approximation. Therefore, for the numerical experiments in Sec.~\ref{sec:numerical_experiments}, we used the microcanonical SBM and set $c_1=10$, so that the effect of the approximation can be ignored.

\section{Relationship with the mixed-membership SBM}
\label{sec:relationship_with_MMSBM}
Mixed-membership stochastic block model (MMSBM) \cite{airoldi2008mixed} is a popular random graph model that considers an overlapping structure. In this section, we discuss the relationship between our overlapping SBM and the MMSBM. We define a membership vector of node $i$ as $\boldsymbol{\pi}_i=[\pi_{ik}]$ $(k\in\{1,\dots,K\})$, $\sum_{k=1}^K \pi_{ik}=1$, $0<\pi_{ik}\leq 1$. 
That is, ${\pi}_{ik}$ represents the probability that node $i$ is assigned to block $k$. In the MMSBM, the edge generation probability of a pair of nodes $(i,j)$ is expressed as
\begin{equation}
    P(A_{ij}=1|\boldsymbol{\rho}, \boldsymbol{\pi}_i,\boldsymbol{\pi}_j)=\boldsymbol{\pi}_i^{\top}\boldsymbol{\rho}\boldsymbol{\pi}_j.
    \label{eq:edge_generation_prob_pi_rho_pi}
\end{equation}
To see the correspondence to our overlapping SBM, we consider a two-block MMSBM, and exclusive node sets $V_1$, $V_2$, and $V_3$, where $V_1$ and $V_3$ represent community blocks and $V_2$ represents the overlapping block. For example, let us consider the following parameterization of $\boldsymbol{\pi}_i$.
\begin{equation}
    \boldsymbol{\pi}_i = \begin{cases}
    (1,0)^{\top} & (i\in V_1) \\
    (1/2,1/2)^{\top} & (i\in V_2) \\
    (0,1)^{\top} & (i\in V_3).
  \end{cases}
  \label{eq:mmsmb_mem_v_pi}
\end{equation}
We consider the same parameterization as Eq.~(\ref{eq:non_overlap_can_affinity_matrix}) for the affinity matrix $\boldsymbol{\rho}$. By inserting Eqs.~(\ref{eq:non_overlap_can_affinity_matrix}) and (\ref{eq:mmsmb_mem_v_pi}) into Eq.~(\ref{eq:edge_generation_prob_pi_rho_pi}), we obtain the edge generation probability matrix
\begin{align}
    \begin{pmatrix}
    \rho_{\text{in}} & \frac{\rho_{\text{in}}+\rho_{\text{out}}}{2} & \rho_{\text{out}} \\
    \frac{\rho_{\text{in}}+\rho_{\text{out}}}{2} & 
    \frac{\rho_{\text{in}}+\rho_{\text{out}}}{2} & \frac{\rho_{\text{in}}+\rho_{\text{out}}}{2} \\
    \rho_{\text{out}} & \frac{\rho_{\text{in}}+\rho_{\text{out}}}{2} & \rho_{\text{in}}
    \end{pmatrix}.
    \label{eq:edge_generation_prob_soft}
\end{align}
This equation never coincides with Eq.~(\ref{eq:can_affinity_matrix}). In fact, one can easily confirm that the MMSBM does not coincide with our overlapping SBM for arbitrary choices of $\boldsymbol{\pi}_i$ in Eq.~(\ref{eq:mmsmb_mem_v_pi}).

It is interesting to consider a variant of the standard MMSBM. We define a membership vector of node $i$ as an unnormalized propensity vector $\boldsymbol{g}_i=[g_{ik}]$ ($k\in\{1,\dots,K\}$), $g_{ik}\geq0$. Similarly to the standard MMSBM, the edge generation probability of a pair of nodes $(i,j)$ is expressed as
\begin{equation}
    P(A_{ij}=1|\boldsymbol{\rho}, \boldsymbol{g}_i,\boldsymbol{g}_j)=\boldsymbol{g}_i^{\top}\boldsymbol{\rho}\boldsymbol{g}_j.
    \label{eq:edge_generation_prob_g_rho_g}
\end{equation}
Again, we consider the case of $K=2$ and the following parameterization of $\boldsymbol{g}_i$.
\begin{equation}
    \boldsymbol{g}_i = \begin{cases}
    (1,0)^{\top} & (i\in V_1) \\
    (\frac{1}{1+\epsilon},\frac{1}{1+\epsilon})^{\top} & (i\in V_2) \\
    (0,1)^{\top} & (i\in V_3).
  \end{cases}
  \label{eq:mmsmb_mem_v_g}
\end{equation}
Here, the labels of the two community blocks are exchangeable because of the permutation symmetry.
By inserting Eqs.~(\ref{eq:non_overlap_can_affinity_matrix}) and (\ref{eq:mmsmb_mem_v_g}) into Eq.~(\ref{eq:edge_generation_prob_g_rho_g}), we obtain the edge generation probability matrix
\begin{equation}
    \begin{pmatrix}
    \rho_{\text{in}} & \rho_{\text{in}} & \rho_{\text{out}} \\
    \rho_{\text{in}} & 
    \frac{2}{1+\epsilon}\rho_{\text{in}} & \rho_{\text{in}} \\
    \rho_{\text{out}} & \rho_{\text{in}} & \rho_{\text{in}}
    \end{pmatrix}.
    \label{eq:edge_generation_prob_hard}
\end{equation}
Equation~(\ref{eq:edge_generation_prob_hard}) becomes identical to Eq.~(\ref{eq:can_affinity_matrix}) when $\sigma=2/(1+\epsilon)$.
In fact, one can confirm that the parameterization of $\boldsymbol{\pi}_i$ in Eq.~(\ref{eq:mmsmb_mem_v_g}) is the only nontrivial choice that achieves the equivalence to our overlapping SBM.
Therefore, this generalized MMSBM and our overlapping SBM share the same model space in the range of $1\le\sigma\le2$.

\nocite{*}

\end{document}